# Exciton-resonant detection of high-frequency surface acoustic waves from subwavelength metal gratings

Olga Ken[1,*], Dmytro Horiachyi[1], Ilya Akimov[1], Vladimir Korenev[1], Vitalyi Gusev[2], Leonid Litvin[3], Michael Kahl[3], Arne Ludwig[4], Nikolai Spitzer[4], Andreas D. Wieck[4], and Manfred Bayer[1,5]

[1] *Experimentelle Physik 2, Technische Universität Dortmund, 44221 Dortmund, Germany*
[2] *Laboratoire d'Acoustique de l'Université du Mans, UMR CNRS 6613, Le Mans Université, 72085 Le Mans, France*
[3] *Raith GmbH, 44263 Dortmund, Germany*
[4] *Lehrstuhl für Angewandte Festkörperphysik, Ruhr-Universität Bochum, 44801 Bochum, Germany*
[5] *Research Center FEMS, Technische Universität Dortmund, 44227 Dortmund, Germany*

**ABSTRACT.** We report on all-optical generation and detection of high-frequency (up to ~30 GHz) surface acoustic waves (SAWs) in GaAs/AlGaAs heterostructures with short-period Au gratings on top. A highly sensitive method for SAW detection is demonstrated using a polarization-resolved pump–probe technique that exploits the narrow exciton resonance in high-quality GaAs. The elastic strain of the SAW modulates the exciton energy in the time domain. As a result, even a small deformation produces a noticeable change in the dielectric function at the detection wavelength leading to an order of magnitude increase in the detection sensitivity as compared to off-resonant conditions. A theoretical model is developed that considers two detection schemes: one accounting for probe light diffraction and one corresponding to a non-diffractive situation.

[*] Contact author: olga.ken@tu-dortmund.de

## I. INTRODUCTION

Recently, there has been growing interest in investigating a new direction of semiconductor physics – chiral phononics – the main goals of which are generation, manipulation and detection of elliptically polarized elastic waves in a solid state, i.e. chiral phonons [1–3]. Similar to photons, elliptically polarized phonons can be characterized by their angular momentum. Such phonons, either acoustic or optical, can exist in systems where either the time-reversal symmetry [4–7] or the spatial inversion symmetry [3,8] is broken. Chiral phonons can be launched on the surface of a non-magnetic semiconductor in the form of Rayleigh surface acoustic waves (SAWs) [9,10], which exhibits an elliptical displacement. This means that a Rayleigh SAW is elliptically polarized and carries the angular momentum, which lies in the plane of the surface, along which the SAW propagates, and is perpendicular to the SAW wave-vector [11].

In view of the vastly expanding interest in magnetization of charge carrier spins by chiral phonons [3,5,12–15], SAWs seem to be promising candidates [11,16]. Here the interaction between SAWs and excitons plays a crucial role and it has been extensively studied in the low-frequency limit: it was shown that in GaAs the piezoelectric field of a SAW can effectively modulate the bandgap, as well as the carrier and spin distribution [17–19]. However, for efficient transfer of angular momentum, it is crucial that the SAW frequency $f_{\mathrm{SAW}}$ is high enough to fulfil the condition: $f_{\mathrm{SAW}} > \tau_{\mathrm{X}}^{-1}, \tau_{\mathrm{sX}}^{-1}$, where $\tau_X$ and $\tau_{sX}$ are the exciton lifetime and the exciton spin relaxation time, respectively. For GaAs, it means that the frequency of SAW should be in the 10–100 GHz range.

For launching SAWs with such frequencies, it is favorable to use methods of the picosecond laser ultrasonics [20,21], i.e. ultrafast strain generation via absorption of pico- or femtosecond laser pulses. SAWs in the sub-GHz range can be launched through the thermoelastic effect produced by absorption of focused light on a metal film deposited on a semiconductor surface [22]. In this case, the SAW wavelength typically lies in the micrometer range. Alternatively, metal nanostructures such as nanorods [23] or gratings [24–27] can be used. For the latter, the SAW wavelength is defined by the grating period, and the generation of GHz-frequency SAWs requires sub-optical-wavelength metal gratings [28,29].

Detection of GHz-frequency SAWs has been realized using time-resolved reflection or transmission pump–probe techniques [24,26] and photoluminescence measurements [30,31]. Detection sensitivity is a crucial aspect that must be carefully considered. For example, interferometric detection schemes [28,29] can significantly enhance sensitivity. Excitonic resonances also provide a highly sensitive probe for acoustic waves, as demonstrated for bulk longitudinal acoustic phonons [32,33]. However, to the best of our knowledge this approach has not yet been applied for detection of high-frequency SAWs.

In this work, we perform all-optical generation and detection of high-frequency (up to ~30 GHz) SAWs on GaAs/AlGaAs heterostructures with sub-optical wavelength gold (Au) gratings (period down to 100 nm). Thermal expansion of the Au stripes upon absorption of femtosecond pump pulses launches both bulk and surface acoustic waves in the heterostructure. We show that the SAWs propagate over distances exceeding 5 μm. Using a polarization-resolved pump-probe technique we present a sensitive method of SAW detection, which exploits the narrow exciton resonance in high-quality GaAs. The elastic strain of the SAW causes modulation of the exciton energy in the time domain. As a result, even small strain of ~$10^{-5}$–$10^{-4}$ magnitude leads to a noticeable change of the dielectric function at the detection wavelength near resonance. This accounts for an order of magnitude increase in the detection sensitivity as compared to detection apart from the resonance. Moreover, instead of conventional measurements of the reflected intensity, we detect the angle of rotation of the reflected probe light polarization caused by the SAW-induced optical anisotropy. To that end, a balanced detection scheme is used, which provides an increase of the signal-to-noise ratio and an even larger enhancement of the sensitivity. We develop a theoretical model which considers possible schemes for optical detection of a sub-optical wavelength SAW and reveals the importance of the metal grating not only in generation of high-frequency SAWs, but also in their detection. These results advance the investigation of angular momentum transfer from elliptically polarized SAWs to charge carriers, which together with previous work on bulk chiral phonons [12–15] pave the way towards phonon-mediated spin orientation.

## II. EXPERIMENTAL DETAILS

We study an AlGaAs/GaAs/AlGaAs heterostructure which is grown on a semi-insulating (100)-oriented GaAs substrate by molecular-beam epitaxy. It consists of a ~1-µm-thick $Al_{0.34}Ga_{0.66}As$ layer, followed by a 50-nm-thick GaAs layer and a 10-nm-thick $Al_{0.34}Ga_{0.66}As$ layer (Fig. 1(a)). On the top a 5-nm-thick GaAs cap layer is deposited in order to prevent the AlGaAs from oxidation. The gold gratings are patterned on the sample surface by means of electron beam lithography and "lift-off" structuring. The grating period varies from 100 to 400 nm, the duty cycle is 50%. The lateral dimensions of each grating are 10×10, 10×20, or 20×50 µm$^2$, the Au thickness is 10 nm. The stripes of the gratings are aligned along the [110] direction of the GaAs. Figure 1(b) shows a scanning electron microscopy (SEM) image of the grating with a period of 120 nm.

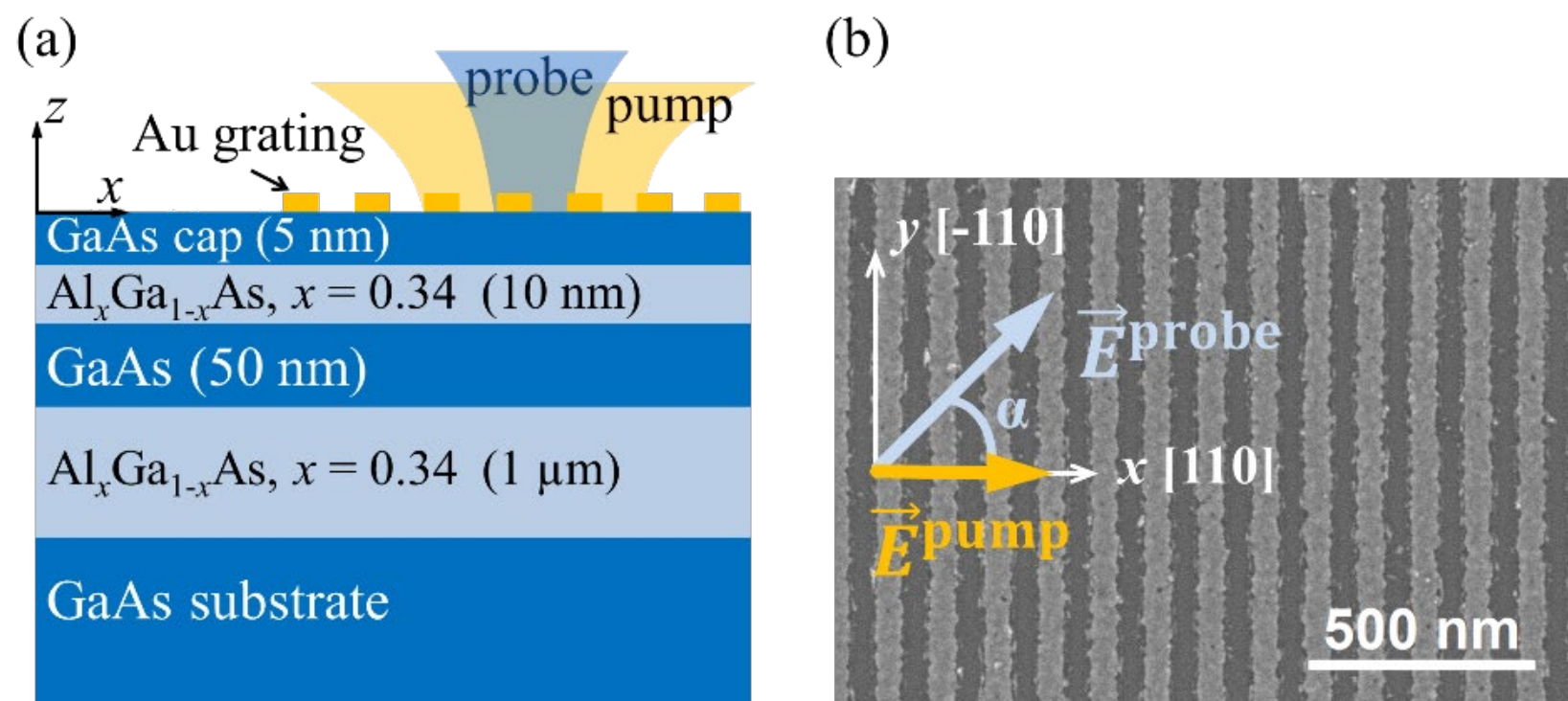

FIG. 1. Sample structure and geometry of the pump-probe experiment. (a) Sketch of the AlGaAs/GaAs/AlGaAs structure with Au grating on top and scheme of the pump-probe experiment. (b) SEM image of the Au grating with a period of 120 nm and a scheme showing directions of the pump and probe linear polarization (*E*-fields).

In order to characterize the structure, the reflectivity and photoluminescence (PL) spectra are measured at a temperature of $T = 5$ K using a spectrometer with a charge-coupled device camera. The PL is excited by a cw Ti:Sapphire laser with a wavelength of 750 nm and the reflectivity spectrum is measured under white light illumination.

Excitation and detection of the acoustic elastic waves are achieved by means of a polarization-sensitive transient pump-probe technique. We use an asynchronous optical sampling (ASOPS) system Gigajet 20/20c from Laser Quantum, which allows one high-speed time delay scanning without a mechanical delay line. It provides ~100-fs pump and probe pulses at a repetition frequency of around $f_0 \approx 1$ GHz, which are locked together with a frequency offset of 20 kHz. As a result, the relative time delay between the pump and probe pulses is repetitively ramped from zero to ~ 1 ns. The pump wavelength is chosen to be 865 nm, which is in the GaAs transparency region, while the probe wavelength is tuned in the range 815–825 nm, i.e. near the exciton resonance in GaAs at $T = 5$ K. Pulse shapers are used for spectral narrowing of the laser pulses, so that the spectral width of the pump is 10 nm, and spectral width of the probe is ~1 nm. The energy densities per pulse are ~100 µJ/cm$^2$ (pump) and ~10 µJ/cm$^2$ (probe). A microscope objective with 20× magnification and a numerical aperture of 0.40 is used for focusing the pump and probe beams on the surface of the sample in spots with diameter of about 4 and 3 µm, correspondingly. While in most experiments the pump and probe beams are focused at nearly normal incidence on the same region of the grating, measurements of the SAW propagation length are performed with the pump and probe spots spatially separated on the grating. The light reflected from the sample is collected with the same microscope objective and the pump wavelength is filtered out by a short-pass spectral filter.

The pump beam is polarized perpendicular to the stripes of the gratings. The probe beam is polarized at an angle $\alpha \approx 45°$ to the grating stripes (see Fig. 1(b)). The angle of the probe beam polarization is chosen from the following symmetry considerations. The presence of the Au grating and the grating-generated SAW set an in-plane anisotropy, giving rise to different reflection and scattering coefficients along and across the preferential axis, i.e. parallel or perpendicular to the grating stripes. The situation is similar to that for a birefringent half-wave plate: if the incident light is polarized at 45° to the fast axes of the waveplate, the rotation of its polarization plane is maximal. The rotation angle ($\theta_R$) of the polarization plane of the reflected probe beam is measured as a function of time delay between the pump and probe pulses. For this, a Wollaston prism in conjunction with a balanced photodetector is used. In the balanced scheme, the change of the polarization rotation angle $\vartheta_R(t)$ is proportional to the normalized differential signal $dU(t)$ of the balanced photodetector: $\vartheta_R = \frac{1}{4}\frac{dU}{U_0} = \frac{I_x - I_y}{4 I_x^0}$. Here $I_{x,y}$ are intensities of $x$ and $y$ polarized components of the reflected probe beam. The balanced scheme allows compensating the signal components (denoted by "0") in the absence of the SAW (when the pump beam is blocked) by rotation of the polarization sensitive beamsplitter, so that $I_x^0 = I_y^0$. The amplified differential signal is sent to a high-speed multichannel digitizer triggered by the ASOPS system.

Employing a balanced photodetector enables effective noise compensation and a substantial improvement in signal-to-noise ratio. Since the two detection channels collect orthogonally polarized components of the same probe beam reflected from the sample surface, even noise arising from mechanical vibrations of the cryostat is largely suppressed. In our setup, which employs a laser with a high pulse repetition rate (~1 GHz), this scheme provides a substantial reduction of background noise. Together with the ASOPS system, which minimizes the impact of slow fluctuations such as laser intensity drifts, this setup allows measurement of the rotation angle down to $10^{-8}$ rad.

All the experiments are carried out in a flow cryostat at a temperature of $T = 5$ K.

## III. EXPERIMENTAL RESULTS

Excitation of the Au grating on top of the sample with femtosecond pump pulses causes its thermal expansion, which launches strain waves. The pump wavelength is chosen in the transparency region of GaAs, so the optical energy is primarily absorbed in the Au grating. When the pump and probe spots overlap on the Au grating, a transient signal is observed, which can be fitted well by an exponential decay superimposed by two damped sine waves, as depicted in Fig. 2(a) for the grating period $a_{\mathrm{gr}} = 160$ nm. The exponentially decaying part of the signal may originate either from transient heating of the Au grating, or from the impurity-band optical transitions under pump excitation. The two damped sine waves correspond to the two dominant frequencies in the Fourier transform (FT) spectra (Fig. 2(b)).

The oscillations starting from the delay of ~220 ps after the pump pulse (green curve in Fig. 2(a)) result from the interference of the probe beam reflected from the stationary interfaces of the sample and the one scattered from the front of the longitudinal acoustic (LA) strain pulse propagating with the sound velocity $v_{\mathrm{LA}}$ perpendicular to the surface into the bulk of the structure. Indeed, the oscillation frequency of $f_{\mathrm{LA}}$ scales with the probe wavelength $\lambda_{\mathrm{pr}}$ in accordance with the interference condition $2nv_{\mathrm{LA}}/f_{LA} = \lambda_{\mathrm{pr}}$, where $n$ is the refractive index. At $\lambda_{\mathrm{pr}} = (819\pm1.5)$ nm the measured $f_{\mathrm{LA}} = (42.5\pm0.7)$ GHz is consistent with the LA velocity along [001] $v_{\mathrm{LA}} = 4.73$ km/s [34] with the refractive index of $n = 3.68\pm0.06$, which is a reasonable value for undoped GaAs [35]. The delay of ~220 ps corresponds to a propagation distance of approximately 1.1 μm, which matches the thickness of the epitaxially grown GaAs/AlGaAs heterostructure on the GaAs substrate. This indicates that signal observed at time delays $t > 220$ ps arises from the LA pulse propagating further into the GaAs substrate. It is worth noting that in the raw signal shown in Fig. 2(a), additional oscillations at ~42 GHz with a much smaller amplitude are visible within the first 220 ps. We attribute these oscillations to the LA-pulse propagating in the AlGaAs/GaAs/AlGaAs epitaxial part of the sample.

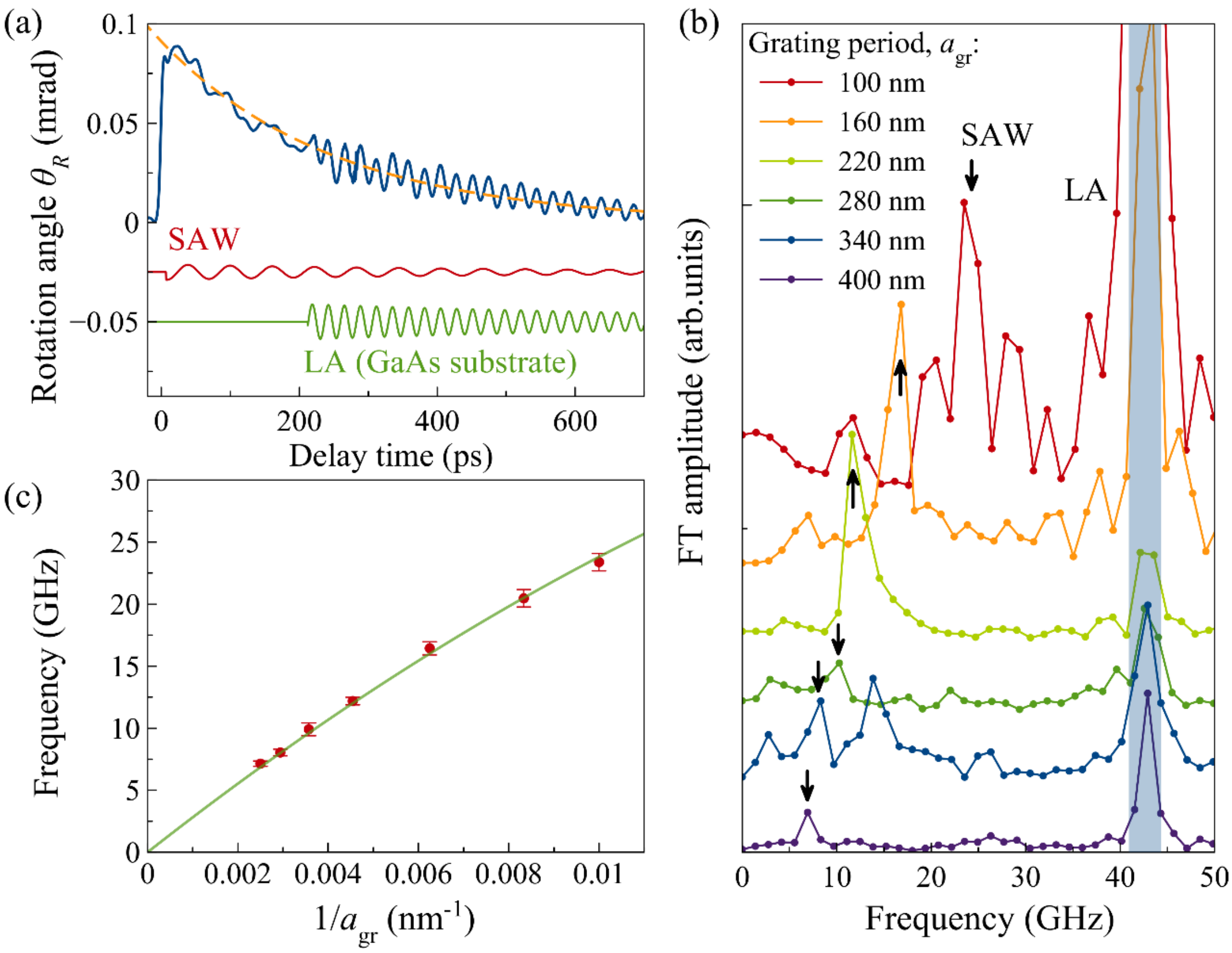

FIG. 2. Generation of coherent SAWs and bulk LA phonons via excitation of an Au grating by fs-pump pulses. (a) Polarization rotation angle $\vartheta_R(t)$ measured for a grating with a period of 160 nm at the probe wavelength of 819 nm, i.e. near the exciton resonance (upper blue curve). The orange dashed line represents an exponential fit of the signal, while the red and green solid curves show sine components related to the SAWs and LA phonons respectively. The curves are shifted vertically for clarity. (b) Fourier transform of the transient signals (after subtraction of the exponential part) for different gratings periods. The spectra are shifted vertically for clarity. The arrows highlight the peaks related to SAWs, while the blue-shaded region indicates the position of the LA peak. (c) Dependence of the SAW frequency on the reverse grating period: experimental data (red dots) and fit according to Eq. (1) (green curve).

Observation of the coherent LA-phonons corresponds to the well-known time-domain Brillouin scattering (TDBS) [21,36] with the only difference that here we measure the rotation of polarization of the acoustically scattered probe beam instead of the change in the reflection intensity, *dR/R*. It is worth noting that coherent LA phonons can also be launched on the free GaAs surface, i.e. without any grating, when the pump wavelength is tuned into the absorption region of GaAs. In this case, LA-related oscillations can be detected via measuring the *dR/R* signal (which is conventional TDBS), but no LA-related oscillations are observed in the $\vartheta_R(t)$ signal due to symmetry reasons. So, for detection of the LA-related oscillations in the polarization rotation signal $\vartheta_R(t)$ a grating is required, which provides the in-plane anisotropy (details are provided in Supplemental Material [37], Section 5, see also references [36,38–49] therein).

In addition to the bulk LA mode, lower-frequency oscillations appear immediately after excitation (red curve in Fig. 2(a)). Their frequency does not depend on the probe wavelength but increases with the decrease in the grating period (Fig. 2(b), (c)). This is a characteristic feature of a SAW, a wavelength of which is determined by the grating period: $\lambda_{\mathrm{SAW}} = a_{\mathrm{gr}}$, and thus its frequency dependence on the reverse grating period is close to linear at least for large grating periods. However, for decreasing grating periods, the SAW velocity is altered from the Rayleigh velocity for a pristine substrate, $v_{\mathrm{R}}$, because of the mass loading effect of the Au gratings. So, the SAW frequency dependence can be better described by accounting for correction $\sim a_{\mathrm{gr}}^{-2}$:

$$f_{\mathrm{SAW}}(a_{\mathrm{gr}}) = v_{\mathrm{R}}\left(1 - \Delta v(a_{\mathrm{gr}})/v_{\mathrm{R}}\right)a_{\mathrm{gr}}^{-1} = v_{\mathrm{R}}\left(a_{\mathrm{gr}}^{-1} - b a_{\mathrm{gr}}^{-2}\right) \quad (1)$$

By taking the Rayleigh velocity for GaAs $v_{\mathrm{R}} = 2.86$ km/s [36,50], a good fit to the experimental data is obtained with $b \approx 17$ nm (the green line in Fig. 2(c)). This means that the deviation from a nondispersive propagation of the SAW is $\Delta v/v_{\mathrm{R}} \cong 0.17$ for shortest grating period ($a_{\mathrm{gr}}$ =100 nm). It is in good agreement with the estimations from Ref. [51], which gives $\Delta v/v_{\mathrm{R}} = 5.11\gamma\, h/a_{\mathrm{gr}} \approx 0.25$ for the Au thickness $h = 10$ nm and the duty cycle $\gamma = 50\%$.

In the FT spectra of some gratings (e.g. with periods 100 and 340 nm) several more strong peaks are observed. These additional peaks can be related, e.g., to the vibrations of the Au stripes [26,52], or to the Brillouin scattering of the probe light diffracted by the grating in high diffraction orders [53].

Our experiments show that the amplitude of the SAW-related oscillations of $\vartheta_R(t)$ strongly depends on the probe wavelength: for different grating periods it reaches maximum when the probe is in resonance with the exciton optical transition in the 50-nm GaAs layer. Figure 3(a) shows an example of this behavior for the grating period $a_{\mathrm{gr}}$ = 220 nm. The amplitude of the SAW-related sine component of $\vartheta_R(t)$ changes by an order of magnitude along the resonance contour reaching its maximum at the probe wavelength of ~819 nm. This value corresponds to the position of the exciton feature in the reflectivity spectrum and the peak in the PL spectrum measured at the free surface of the sample at $T$ = 5 K (Fig. 3(b)). It agrees well with the position of the exciton resonance in bulk GaAs (~818 nm at 2 K). The slight shift, which is within the exciton linewidth, can be attributed to laser-induced heating or to residual strain in the 50-nm GaAs layer.

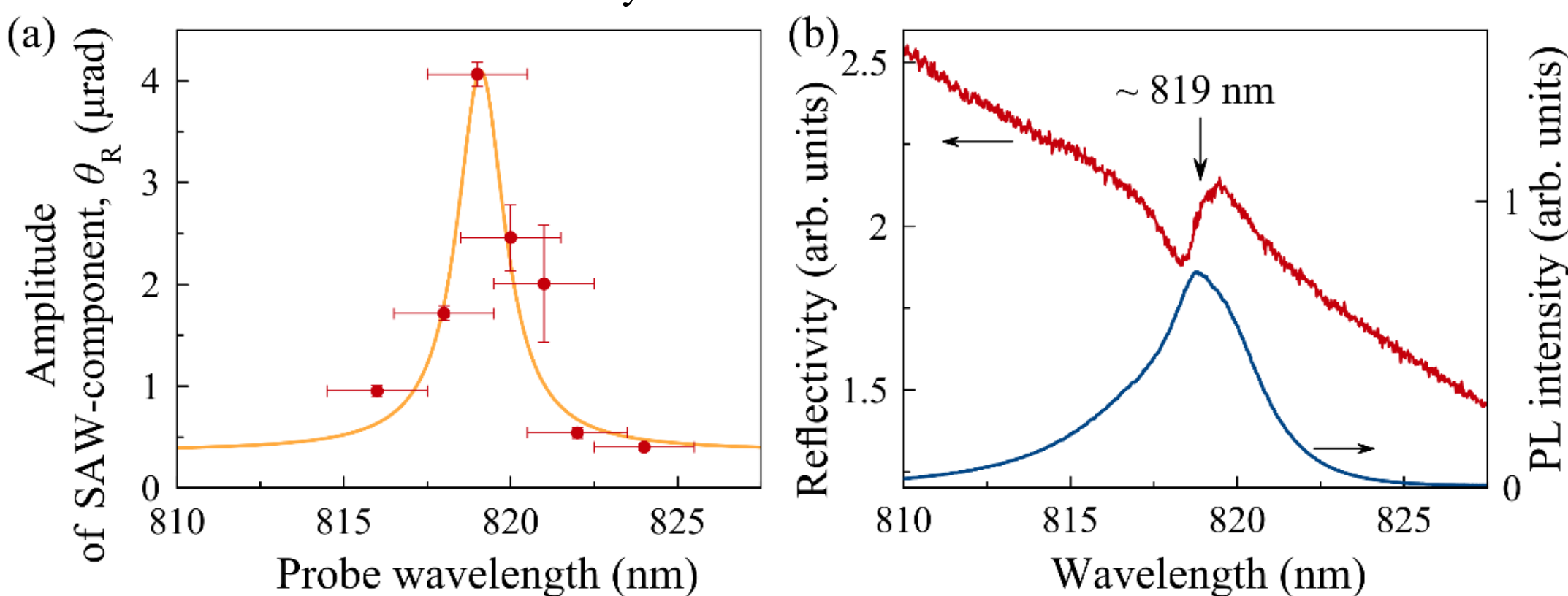

FIG. 3. Exciton resonance in SAW detection. (a) Amplitude of the SAW-related sine component of $\vartheta_R(t)$ as a function of the probe wavelength for the grating period $a_{\mathrm{gr}}$= 220 nm, the horizontal error bars correspond to the probe spectral width. The solid line shows a Lorentzian fit (centered at 819.0 nm with a FWHM = 1.8 nm). (b) Reflectivity (red) and PL spectrum (blue) of the plane surface of the sample (without the Au grating), both showing a feature related to the exciton optical transition in the 50-nm GaAs layer.

Measurements with the pump and probe spots spatially separated from each other at the Au grating, confirm that the SAWs are propagating along the surface mainly in direction perpendicular to the grating stripes. Figure 4(a) shows a $\vartheta_R(t)$-signal transient measured at a grating with the period $a_{\mathrm{gr}}$ = 340 nm when pump and probe spots are displaced from each other by 6 μm across the grating (i.e., perpendicular to the grating stripes, see the inset in Fig. 4(a)). The oscillations related to the SAW are clearly seen in the transient signal, and its FT spectrum shows a pronounced peak at $f_{\mathrm{SAW}}$ = 8.1 GHz (Fig. 4 (a), (b)). Its amplitude changes only slightly with the variation of the separation between the pump and probe spots (Fig. 4(c)). The non-zero amplitude of the LA-related oscillations can result from the partial overlap of the Gaussian wings of the pump and probe beams separated in space. However, from Fig. 4(c) it is obvious that, with increasing pump-to-probe distance, the amplitude of the LA oscillations decreases much steeper than that of the SAW-related signal. It is worth noting, that when the pump and probe spots are separated by 5 μm along the grating stripes, the SAW-related signal is not observed.

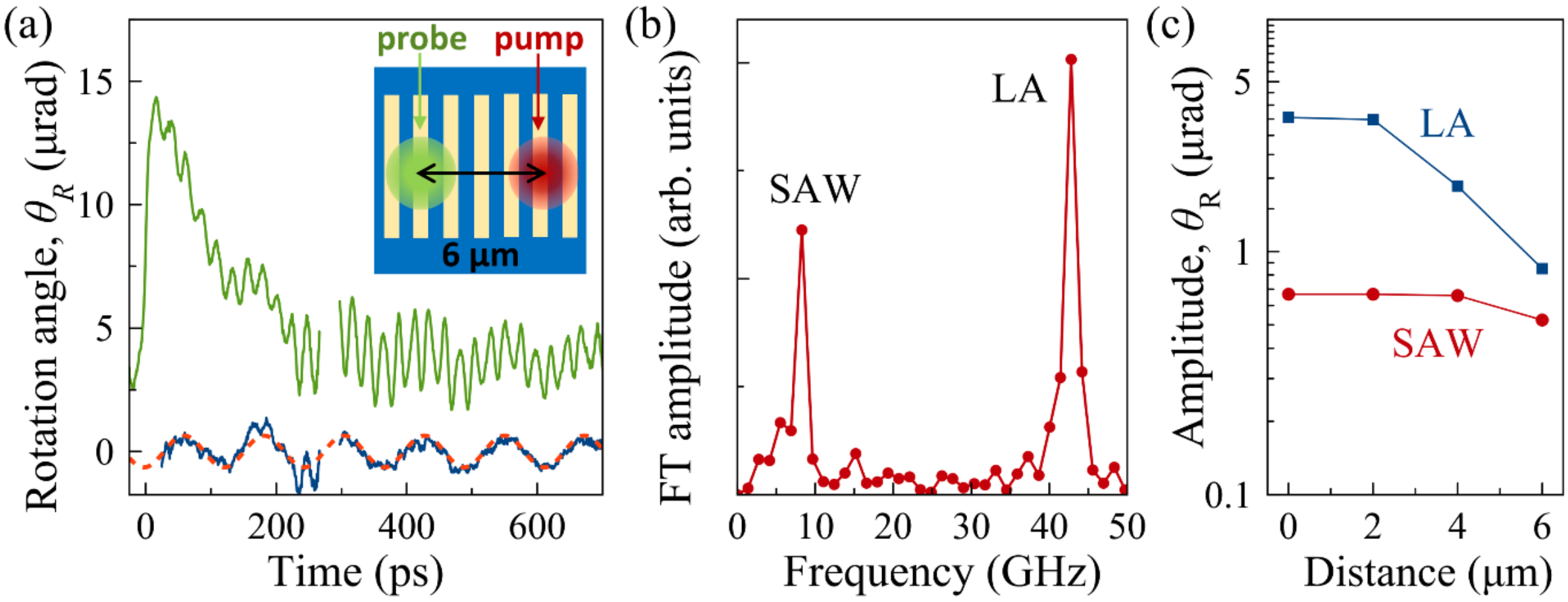

FIG. 4. Experiments with spatially separated pump and probe. (a) Polarization rotation angle $\vartheta_R(t)$ measured for the grating period $a_{\mathrm{gr}}$ = 340 nm, when the pump and probe beams are separated by 6 μm across the grating: raw signal (green line), subtracted SAW contribution (blue line), and its sine fit with a frequency of 8.1 GHz (red dashed line). The inset shows schematic geometry of the experiment. (b) FT amplitude of the $\vartheta_R(t)$ signal with subtracted exponential part. (c) Amplitudes of the LA- and SAW-related oscillation components of $\vartheta_R(t)$ as a function of the distance between the pump and probe spots across the grating ($a_{\mathrm{gr}}$ = 340 nm).

Although the propagation distance of several micrometers observed here is much smaller than those typical of MHz-frequency SAWs on a free surface [54], it is comparable to the results reported for Al gratings on silicon, where similar propagation length was measured for SAWs with frequencies above 10 GHz [26]. SAWs propagating along a loaded surface (i.e., covered by the metal grating) are dispersive and experience substantially stronger damping than free-surface SAWs, which constitutes the dominant loss mechanism in the 1–100 GHz range [55]. Moreover, prior studies of GHz-frequency SAWs indicate that the SAW lifetime – and hence the propagation length – decreases with increasing of the SAW frequency [56,57]. The metal-grating duty cycle further affects damping, with a 50% duty cycle used in our structures yielding the maximum attenuation [24].

## IV. MODEL AND DISCUSSION

In this section we consider the possible schemes for optical detection of sub-optical wavelength SAWs that are relevant to our experiments, and model the measured signal in polarization rotation $S \sim \vartheta_R$ and its dependence on the grating period.

The probe beam, $\vec{E}^0 \equiv \vec{E}^{\text{probe}}$, is set to be at normal incidence on the sample (along the O*z* axis) and is linearly polarized at an angle $\alpha \approx 45°$ to the Au stripes, i.e. approximately along the [100] direction of the GaAs (see Fig. 1(a)). In this geometry, the signal related to the SAWs is reaching its maximum value. The amplitude of the probe light reflected from the stationary surfaces of the sample is $E_i^r = r_i E_i^0$, where $i = x, y$ stands for the projections on the coordinate axes O*x* || [110], O*y* || [-110]. We consider the anisotropy of the reflection coefficient from the grating $r_x \neq r_y$, as well as its grating-induced spatial modulation: $r_i = \overline{r_i} + \widetilde{r_i}\cos(2\pi x/a_{\text{gr}})$. Here and below, the "tilde" and "bar" symbols indicate the laterally modulated and laterally homogeneous parts, respectively. The pump pulse excites a standing surface acoustic wave, which comprises two counter-propagating waves with wavevector components along O*x* equal to the wavevector of the grating $k_g = \pm 2\pi/a_{\text{gr}}$. The amplitude of the probe light scattered by the SAW, $E_i^s$, is determined by the modification of the GaAs dielectric tensor in the acoustic field, which for our case of cubic crystal can be written as $\Delta\varepsilon_i = \overline{\Delta\varepsilon_i} + \widetilde{\Delta\varepsilon_i}\cos(k_g x) \sim p_{ij}\eta_j$, where $p_{ij}$ is the GaAs photo-elastic tensor and $\eta_i$ are the strain components of the SAW.

For detection of the coherent acoustic waves at GHz frequencies, generated by ultrashort pump laser pulses, the weak electric field of the probe light scattered by the acoustic wave, $\vec{E}^s$, should interfere at the photodetector with the much stronger electric field of the probe light reflected by the stationary surfaces of the sample, $\vec{E}^r$ [21,46]. So the resulting optical intensity is $I = \left|\vec{E}^r + \vec{E}^s\right|^2 \cong \left|\vec{E}^r\right|^2 + 2Re\left[\left(\vec{E}^r\right)^* \vec{E}^s\right]$, where we neglect the smallest term which is the intensity of the scattered light. To measure the polarization rotation of the acoustically scattered probe light, we split the two orthogonal components of the probe electric field reflected from the sample and measure their intensities, $I_x$ and $I_y$, separately by the two photodetectors. The polarization rotation angle $\theta_R(t)$ is proportional to the difference between the two signals of photodetectors $dU$. The scheme is balanced in the absence of coherent acoustic waves, i.e., when blocking the pump laser, so that $I_x^0 = I_y^0 \cong r_0^2 E_0^2/2$, where $E_0^2 = \left|\vec{E}^r\right|^2$, and $r_0$ is the reflection coefficient of the bare GaAs/AlGaAs structure. The differential signal is then normalized on $U_0 \propto I_x^0$, so that the measured signal is:

$$S \sim \vartheta_R \sim \frac{dU}{U_0} = \frac{I_x - I_y}{2I_0} \approx \frac{Re\left[(E_x^r)^* E_x^s - (E_y^r)^* E_y^s\right]}{r_0^2 E_0^2}. \tag{2}$$

Periodic modulation of the sample surface along O*x* due to the Au grating causes diffraction of the probe light. The wavelength of the probe light inside the structure is $\lambda_{\text{pr}}/n \approx 230$ nm, where $n \approx 3.68$ is the GaAs refractive index determined above. The period of the Au gratings varies from 100 to 400 nm, so that the short-period gratings ($a_{\text{gr}} = 100$, 120, 160 nm) are sub-optical. The same holds for the SAWs, whose elastic strain gives rise to the spatial modulation of the optical properties of the GaAs/AlGaAs structure with the same period ($\lambda_{\text{SAW}} = a_{\text{gr}}$). For sub-optical gratings, only the probe light reflected and transmitted in the zeroth diffraction order is propagating into the far-field, while that reflected or transmitted in other diffraction orders is evanescent.

In general, there are numerous paths for the probe light to reach the photodetector after two diffractions by the grating and one diffraction by the SAW. However, for the 10-nm-thick gold grating, the diffraction efficiency is weak and the dominant contribution to the detected signal comes from the acoustically scattered probe light, which is either never diffracted or is diffracted by the grating only once (for details see Supplemental Material, Section 5 [37]). Figure 5 presents three possible paths for the SAW detection relevant for our experiment. The first two paths (Fig. 5 (a), (b)) include (though in different order) one diffraction by the grating and one – by the SAW. It is worth noting that we are considering only the light diffracted into the ±1 diffraction orders, because with the duty cycle of the gratings of 50%, the contributions from higher diffraction orders are negligible. The path without any diffraction (Fig. 5(c)) is possible, because the SAW in the spatially periodic samples is a generalized Rayleigh SAW (gRSAW), which contains a spatially unmodulated (homogeneous) component oscillating at the SAW frequency (see Supplemental Material, Section 3 [37]).

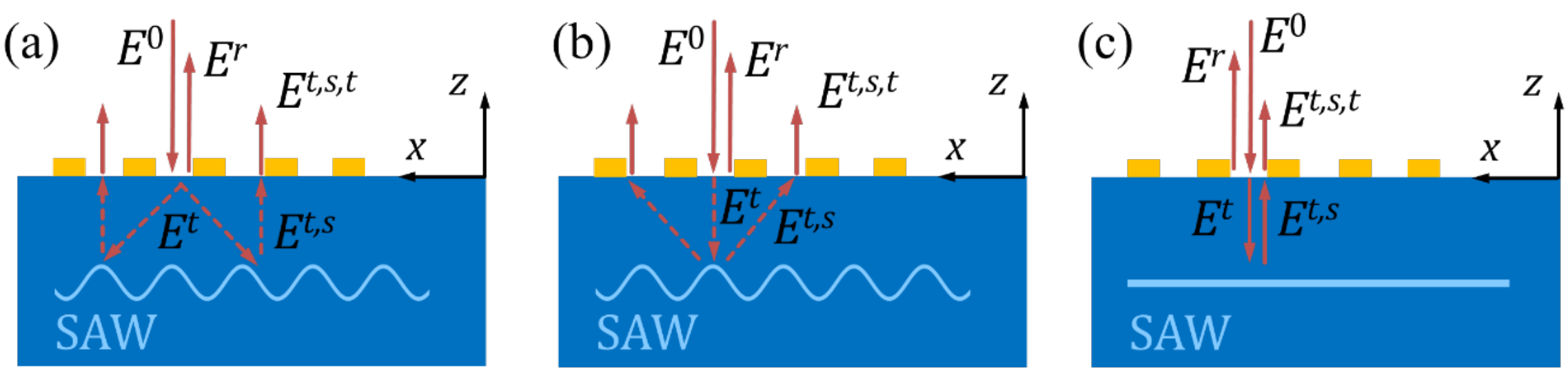

FIG. 5. Detection schemes. Schematic diagram of the probe light paths showing probe beam incident on the sample ($E^0$), reflected from the sample surface ($E^r$), transmitted though the Au grating into the sample ($E^t$), then scattered by the SAW ($E^{t,s}$) and transmitted again through the grating ($E^{t,s,t}$). (*a*) The incident probe beam is diffracted by the grating and the SAW and then transmitted into the far field without diffraction. (*b*) The incident beam is transmitted without diffraction, diffracted by the SAW and then by the grating into the far field. (*c*) The beam is transmitted and scattered without diffraction.

The first two, diffraction-based, paths (Fig. 5(a), (b)) give the following contribution to the total signal $S \sim \vartheta_R$:

$$S_d \sim \mathrm{Im}\{c\tilde{t}_+ J_{d-}\}, \tag{3}$$

where $\tilde{t}_+$ accounts for the isotropic part of the transmission coefficient, which is modulated along the O$x$ axis due to the presence of the grating. The anisotropic part of the diffraction-related scattering integral, $J_{d-} \sim \frac{\Delta\tilde{\varepsilon}_x - \Delta\tilde{\varepsilon}_y}{\varepsilon}$, accounts for the rotation of the probe electric field due to the anisotropy of the permittivity response to the strain: $(\Delta\tilde{\varepsilon}_x - \Delta\tilde{\varepsilon}_y) \sim p'_{44}\tilde{\eta}_x$, where $p'_{ij}$ represent the components of the photo-elastic tensor in the cubic crystal axes of GaAs. Thus, spatial modulation of the optical transmission induced by presence of the Au grating provides the opportunity to detect the spatially modulated component of the SAW, $\tilde{\eta}_x$, which rotates the probe electric field, by insuring that the acoustically scattered light finally propagates outside of the sample towards the detector (even if at the intermediate stages of the optical path it is evanescent or propagating in non-zeroth diffraction order). In the coefficient $c \equiv \frac{1}{2}\left(1 + \frac{1}{\sqrt{1-(k_g/k_n)^2}}\right)$, the first, unity, term accounts for the diffraction path depicted on Fig. 5(a) and the second term, $\left(1 - (k_g/k_n)^2\right)^{-1/2}$, comes from the diffraction path in Fig. 5(b). Here $k_g = 2\pi/a_{\mathrm{gr}}$ is the grating wave number and $k_n = 2\pi n/\lambda_{\mathrm{pr}}$ is the wavevector of the probe light in the media (GaAs). Owing to this coefficient, the diffraction-based part of the signal, $|S_d|$, reaches its maximum for $a_{\mathrm{gr}} \approx \lambda_{\mathrm{pr}}/n$, which for the probe wavelength near the exciton resonance (~818 nm) is best fulfilled for $a_{\mathrm{gr}}$ = 220 nm. In this case, i.e. when the probe wavelength in the medium is comparable with the grating period, the imaginary part of the complex refractive index $n$, i.e., the probe light absorption has significant influence on the signal amplitude (for details see Supplemental Material, Section 5 [37]).

The contribution to the total signal corresponding to the non-diffraction path (Fig. 5(c)) is determined by the imaginary part of the product of the anisotropic part of the reflection tensor $\bar{r}_-$ and the isotropic part of the scattering integral, $J_{p+}$:

$$S_p \sim \mathrm{Im}\{\bar{r}_- J_{p+}\}. \tag{4}$$

The scattering integral $J_{p+} \sim \frac{\Delta\bar{\varepsilon}_x + \Delta\bar{\varepsilon}_y}{\varepsilon}$ in Eq. (4) has a form typical for picosecond laser ultrasonics [20,36], where the change in the dielectric tensor $(\Delta\bar{\varepsilon}_x + \Delta\bar{\varepsilon}_y) \sim p'_{12}\,\bar{\eta}_z$ is determined by the spatially homogeneous component of the strain $\bar{\eta}_z$ induced by the gRSAW acoustic field. From symmetry considerations it follows that this strain component does not rotate the polarization plane of the scattered probe light. Therefore, one needs the grating-induced anisotropy of the optical properties of the sample surface, $\bar{r}_- \neq 0$, (primarily caused by the plasmonic effect [58]), to detect the polarization rotation signal. Moreover, it is also due to this anisotropy that we measure the TDBS signal in polarization rotation from the laterally homogeneous LA waves (see Fig. 2(a), (b)).

The theoretical considerations given above suggest that the metal grating plays an essential role in the detection process, and that without it, the sub-optical SAWs in our experiments are not detectable. The role of the metal grating is twofold. First, the grating provides the diffraction path that allows for detection of the periodically modulated anisotropic component of the SAW. Second, the grating-induced anisotropy provides the opportunity to detect the laterally homogeneous component of the SAW acoustic field, which by itself does not rotate the polarization plane of the probe light.

The GaAs dielectric function changes sharply along the exciton resonance contour. The elastic strain induced by the SAW shifts the resonance frequency of exciton, so that the change in the dielectric function is maximal in the vicinity of the resonance. This leads to an enhancement of the SAW-related signal amplitude by an order of magnitude (see Fig. 3(a)). From these data we estimate the amplification of the photo-elastic constants at the exciton resonance as $p'_{12}(\mathrm{X})/p'_{12} \approx p'_{44}(\mathrm{X})/p'_{44} \approx 10 - 50$ (see Supplemental Material, Section 5 [37] for details). Thus, the photo-elastic tensor components show a huge maximum inside the GaAs layer due to the exciton resonance. It is worth noting, however, that the photo-elastic constants estimated here are much smaller than those obtained for exciton-polaritons in specially designed structures ($3\cdot10^2$ [32,33], up to $\sim10^5$ [59]).

Figure 6(a) shows the dependence of the amplitudes of the calculated signal components $S_d$ and $S_p$ as function of the grating period. As anticipated, the diffraction-related contribution $|S_d|$ has a pronounced maximum in the vicinity of $a_{\mathrm{gr}}$=220 nm and strongly decreases for shorter periods. Indeed, for the sub-optical gratings (100−160 nm), the diffracted probe light is evanescent and its penetration length $l_p \approx \lambda_{\mathrm{pr}} \Big/ \left( 2\pi n \sqrt{\left(\frac{\lambda_{\mathrm{pr}}/n}{a_{\mathrm{gr}}/2}\right)^2 - 1} \right) \approx 20$ nm does not provide good overlap with the GaAs excitonic layer (which is located at 15−65 nm below the sample surface, see Fig. 1(a)), and thus the exciton-induced enhancement is suppressed. For the sub-optical wavelength gratings ($a_{\mathrm{gr}} < \lambda_{\mathrm{pr}}/n \approx 230$ nm) the main contribution to the total signal should come from the non-diffraction paths, $|S_p|$, which shows a monotonic growth with decreasing grating period. The physical origin of this increase lies in the growth of the homogeneous component of the gRSAW as the grating period decreases. When the period becomes smaller, both the wavelength and the penetration depth of the gRSAW decrease, while the mass loading from the metal stripes within the shrinking volume (which guides the gRSAW beneath the surface) remains unchanged. As a result, the periodic modulation of the sample parameters — which is the main reason for the appearance of the laterally homogeneous component — becomes progressively stronger. The magnitude of the homogeneous component of the SAW, relative to its modulated component, therefore increases proportionally to the modulation (see Section 3 of the Supplemental Material [37]).

The calculated total signal amplitude $|S| = |S_p + S_d|$ is represented in Fig. 6(a) (green solid curve). Its overall behavior showing increase for the short-period gratings is similar to the experimental dependence of the polarization rotation angle $\vartheta_R(a_{\mathrm{gr}})$ measured at the exciton resonance and presented in Fig. 6(b), except that the latter demonstrates no indication of a maximum. So, either the steps in the grating periods in the investigated sample are too large for revealing a peak in the vicinity of $a_{\mathrm{gr}}$~220 nm, which is predicted by theory, or for some reason the detection path relying on diffraction is less important in comparison with the non-diffractive one even for the larger period gratings.

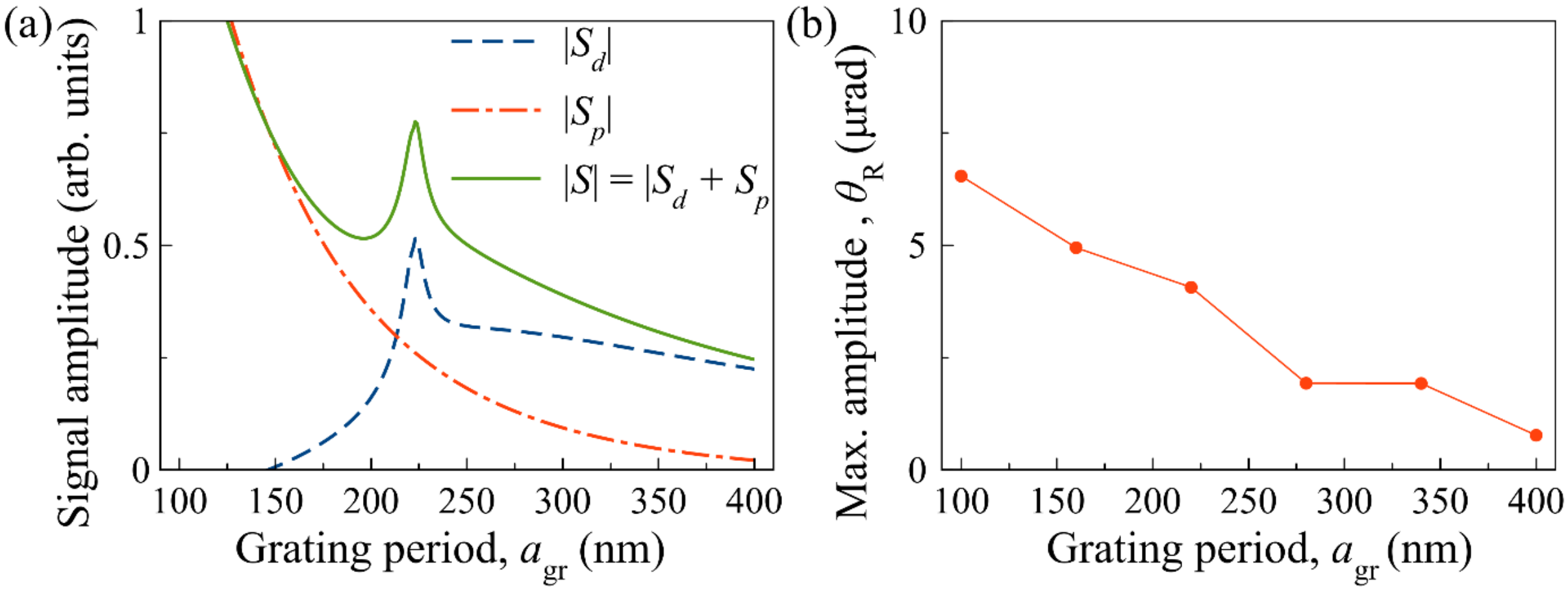

FIG. 6. Amplitude of the SAW-related signal as a function of the grating period. (a) Calculated amplitudes of diffraction-based $|S_d|$ (blue dashed curve) and non-diffraction $|S_p|$ (red dash-dot curve) contributions to the total SAW-related signal $|S| = |S_p + S_d|$ (green solid curve) as a function of the grating period, $n = 3.67 - i0.08$ [47] and $p'_{12}(X)/p'_{12} = p'_{44}(X)/p'_{44} = 10$. (b) Amplitude of the SAW-related signal (rotation angle) measured in the exciton resonance as a function of the grating period.

The growth of the $S_p$ amplitude (and thus the total signal $|S|$) with decreasing $a_{\mathrm{gr}}$ becomes less pronounced when taking into account that the anisotropy of the reflection coefficient, $\bar{r}_-$, itself depends on the grating period. From the measured reflectivity spectra (see Supplemental Material, Section 6 [37]), we find that $|\bar{r}_-|$ reaches its maximum for $a_{\mathrm{gr}}$~160 nm and decreases for other periods, which correspondingly reduces the $S_p$ amplitude. This can potentially improve the agreement between the calculated signal and the experimental data for the sub-optical wavelength gratings ($a_{\mathrm{gr}} < 220$ nm).

## V. CONCLUSIONS

In conclusion, we have demonstrated the excitation of surface acoustic waves with frequencies up to ~30 GHz using ultrafast laser pulses and sub-optical-wavelength Au gratings fabricated on GaAs/AlGaAs heterostructures. Thermal expansion of the Au stripes upon absorption of the pump pulses launches both coherent bulk LA phonons and SAWs in the heterostructure, which are detected simultaneously in the time-domain. The generated SAWs with frequencies around 10 GHz propagate over several micrometers, demonstrating the efficient generation and sustained propagation of high-frequency acoustic modes in the studied structures.

We show that for the detection of sub-optical wavelength (GHz-frequency) SAWs, the metal grating plays a crucial role, and without the periodic modulation of the optical properties introduced by the grating, sub-optical SAWs cannot be detected. The grating-induced in transmission and reflection enables detection of the laterally homogeneous component of the generalized Rayleigh SAW, which by itself does not rotate the probe light polarization. The grating also facilitates detection of the periodically modulated component of the SAWs (which is anisotropic and thus rotates the probe light polarization), by ensuring that the acoustically scattered light is coupled out of the sample toward the photodetector, even though at intermediate stages the light may be evanescent or confined to higher diffraction orders.

We have demonstrated an order of magnitude amplification of the detection sensitivity near the exciton resonance in GaAs compared to off-resonant conditions and estimated a 10−50-fold increase of the photo-elastic constants. This enhancement is attributed to the strong SAW-induced modulation of the GaAs dielectric function near the exciton resonance. Under these conditions, the strain amplitude of ~$10^{-5}$–$10^{-4}$ magnitude, typical for picosecond acoustics, results in a measured rotation angle of $\vartheta_R \sim 10$ μrad.

Finally, the polarization-sensitive pump-probe detection scheme provides a substantial reduction in background noise. In this configuration, the two detection channels of the balanced photodetector collect orthogonally polarized components of the same probe beam reflected from the sample surface, effectively suppressing noise associated with mechanical vibrations of the cryostat. Combined with the ASOPS system, which minimizes the impact of slow fluctuations such as laser intensity drifts, this setup enables the measurement polarization rotation angles down to ~$10^{-8}$ rad.

**ACKNOWLEDGEMENTS**

We thank Serhii Kukhtaruk for initial calculations that provided the start of this woYakovlev for comments and critical reading of the manuscript. O.K., D.H. and V.K. acknowledge financial support from the Deutsche Forschungsgemeinschaft (Projects No. 529684269, No. 53698750, and No.524671439).

**SUPPLEMENTAL MATERIAL**

**for “Exciton-resonant detection of high-frequency surface acoustic waves from subwavelength metal gratings”**

Olga Ken[1], Dmytro Horiachyi[1], Ilya Akimov[1], Vladimir Korenev[1], Vitalyi Gusev[2], Leonid Litvin[3], Michael Kahl[3], Arne Ludwig[4], Nikolai Spitzer[4], Andreas D. Wieck[4], and Manfred Bayer[1,5]

[1] *Experimentelle Physik 2, Technische Universität Dortmund, 44221 Dortmund, Germany*
[2] *Laboratoire d’Acoustique de l’Université du Mans, UMR CNRS 6613, Le Mans Université, 72085 Le Mans, France*
[3] *Raith GmbH, 44263 Dortmund, Germany*
[4] *Lehrstuhl für Angewandte Festkörperphysik, Ruhr-Universität Bochum, 44801 Bochum, Germany*
[5] *Research Center FEMS, Technische Universität Dortmund, 44227 Dortmund, Germany*

## 1. Sample fabrication

The AlGaAs/GaAs/AlGaAs heterostructure was grown on a semi-insulating (100)-oriented GaAs substrate by molecular-beam epitaxy. After thermal oxide removal, we initialize the growth with a GaAs buffer, a short-period superlattice from AlAs and GaAs, and continue, after another GaAs buffer with a ~1-µm-thick $Al_{0.34}Ga_{0.66}As$ layer, followed by a quasi-bulk (50 nm) GaAs layer and a 10-nm-thick $Al_{0.34}Ga_{0.66}As$ layer. On top of the AlGaAs, a 5-nm-thick GaAs cap layer is deposited in order to prevent the Al from oxidation.

Metal gratings on top of the GaAs/AlGaAs heterostructure were fabricated by electron beam lithography and “lift-off” patterning. The lithography was performed using a Raith VOYAGER system operated at 50 kV acceleration voltage using a beam current of 400 pA. A resist of type ZEP-520A was spun 100 nm thick on the wafer surface. Dip development was done in n-Amylacetat for 30 s at 22°C. A 10 nm gold layer was evaporated after deposition of 2 nm of titanium which is necessary for securing adhesion and layer uniformity. The grating sizes were 10×10, 10×20, and 50×20 µm$^2$. We fabricated gratings of different periods such as 100, 120, 160, 220, 280, 340, and 400 nm, with a duty cycle of 50%. To provide line width uniformity over the grating size, proximity effect correction was applied. Still at the edge of a 120 nm grating, for example, the stripe width was 50 nm instead of the targeted 60 nm, i.e., 10 nm less than the nominal width. We explain this by lateral development which results in a reduced line width in the stripe along the grating edge being nearly 3 micrometers wide. In this region the duty cycle changed monotonically from 50% to 42% towards the edge for the period of 120 nm.

## 2. Generation of SAWs by metal grating

The theory of Rayleigh type SAWs by laser-induced stresses distributed in the volume is well developed [1, 2]. It is straightforward to use this general theory. However, we have significantly simplified it by providing estimations using the formulas for depth layered samples [3], which for our particular experimental setting (deposition of a semi-transparent gold layer on the surface of a substrate transparent at the pump wavelength) is justified as the processes of heat conduction from the laser-heated gold into the GaAs substrate plays a negligible role, at least in our case of GHz-frequency SAWs.

[1] Contact author: olga.ken@tu-dortmund.de

The generation of SAWs arises from the thermoelastic stresses in the 10-nm-thick gold layer homogeneously heated by the absorption of a fs-pump pulse. Its temperature rise can be estimated by introducing the absorption $A$ of the gold layer [4]. The structuring of our sample was accounted for by introducing a lateral periodicity of the photo-induced stresses in the gold layer. For the estimates, we have applied the known thermoelastic parameters of bulk gold and the elastic velocities of bulk acoustic waves in elastically anisotropic GaAs, averaged over the directions relevant to our sample orientation. The derived formula for quantitatively estimating the strain component $\eta_x$ along the O$x$ axis can be presented in the following form:

$$\eta_x \sim \left(\frac{f_{\mathrm{SAW}}}{10\mathrm{GHz}}\right) \frac{A \cdot F_L}{\mathrm{nJ}/(\mu\mathrm{m})^2},$$

where $f_{\mathrm{SAW}}$ and $F_L$ denote the SAW frequency and the pump laser energy density, respectively, while $A \approx 0.1$ for a 10-nm gold film at the pump wavelength around 865 nm [4]. The estimate suggests an increase of the SAW amplitude proportional to its frequency, $f_{\mathrm{SAW}}$, i.e., proportional to the inverse period of the metal grating. The amplitudes of the other strain component of the SAW as well as the strain depth distributions, are well documented in the literature (see, for example, [5]). In our case, we arrive at strain values of $\sim 10^{-5}$–$10^{-4}$, typical for picosecond acoustics.

### 3. Estimates of SAW components ratio

Here and further for the theoretical considerations, we will use the $x_1$, $x_2$, $x_3$ notations for the $x$, $y$, $z$ axes used in the main text (and depicted in Fig.1).

Our goal is to estimate the ratio of the amplitudes of the laterally homogeneous and laterally oscillating components of the SAWs, as required for revealing the dominant path of their detection, as described in Supplemental material, Section 5. For the estimates, we will apply here and in the upcoming section the principle of splitting the averaged (laterally homogeneous) and the oscillating components. We consider that the SAW propagates along the O$x_1$ axis across the periodic Au grating with the width of the Au stripes of $d = a_{\mathrm{gr}}/2$, the thickness of $h$ =10 nm and wave number: $k_g = 2\pi/a_{\mathrm{gr}}$) on the GaAs surface. Such a SAW (namely, a generalized Reyleigh SAW, gRSAW) is characterized by the velocities $v_R(0)$ and $v_R(h)$ on the bare GaAs surface and under the gold stripes, respectively. So, the velocity is rectangular modulated along the propagation direction of the gRSAW with the modulation amplitude $\Delta v_R \equiv v_R(0) - v_R(h)$. We present this modulation as a modification of the averaged velocity and the oscillating component with zero average:

$$v_R(x_1) = v_R(0) + \overline{\Delta v_R} + \widetilde{\Delta v_R} = v_R(0) - \frac{\Delta v_R}{2} - \frac{2\Delta v_R}{\pi}\cos(k_g x_1) \equiv \overline{v_R} + \widetilde{v_R}. \quad \text{(S3-1)}$$

For compactness of evaluation we retain only the first harmonic of spatial modulation. Note, that the next term in the Fourier expansion of the rectangular spatial distribution with 50% duty cycle is $\sim\cos(3k_g x_1)$ and its amplitude is 3 times smaller than that of the first harmonic.

The equation for the propagation of the $x_3$ displacement component of the gRSAW, $u_3$, can be qualitatively modeled as

$$\frac{\partial^2 u_3}{\partial t^2} - v_R^2(x_1)\frac{\partial^2 u_3}{\partial x_1^2} = 0. \quad \text{(S3-2)}$$

Note, that for more detailed modeling of the gRSAW structure, one should account for the independent spatial modulation of the mass density and of the elastic moduli [6], however for our estimates Eq.(S3-2) is sufficient for revealing the origin of the homogeneous gRSAW component. Our experiments (see Fig. 2(c) of the main text) revealed a small decrease in the frequency of the detected wave, caused by the gold grating deposition, relative to the frequency, which can be estimated rather precisely from the knowledge of the grating period controlling the gRSAW wavelength and the velocity of the SAW at the bare GaAs substrate. The largest deviation from a nondispersive propagation of the gRSAW takes place for the shortest grating periods of $a_{\mathrm{gr}}$=100 nm: $|\Delta f|/f \cong 0.17$. This provides the opportunity to estimate that in Eq.(S3-1)

$$\left|\frac{\Delta v_R}{2}\right|/v_R(0) \cong |\Delta f|/f \cong 0.17\ . \tag{S3-3}$$

The smallness of the velocity deviations provides the opportunity for the following approximation:

$$v_R^2(x_1) \cong \overline{v_R}^2 - 4\overline{v_R}\frac{\Delta v_R}{\pi}\cos(k_g x_1).$$

We substitute this result in Eq.(S3-2) together with the structure of the gRSAW at the cyclic frequency $\omega = 2\pi f$, in which only the fundamental harmonic is retained in the oscillating part, i.e.,

$$u_3 = \overline{u_3} + \widetilde{u_3} \equiv \overline{u_3} - (\widetilde{u_3})_{k_g}\cos(k_g x_1).$$

Then Eq.(S3-2) takes the form:

$$\omega^2\left[\overline{u_3} + (\widetilde{u_3})_{k_g}\cos(k_g x_1)\right] + (k_g)^2\left[\overline{v_R}^2 - 4\overline{v_R}\frac{\Delta v_R}{\pi}\cos(k_g x_1)\right](\widetilde{u_3})_{k_g}\cos(k_g x_1) \approx 0.$$

The laterally averaged part of this equation provides the relation between the averaged and oscillating components of the gRSAW: $\overline{u_3} \cong 2\frac{(k_g)^2}{\omega^2}\overline{v_R}\frac{\Delta v_R}{\pi}(\widetilde{u_3})_{k_g} \cong \frac{2}{\pi}\frac{\Delta v_R}{v_R(0)}(\widetilde{u_3})_{k_g}$. Here the factor 2 has disappeared after averaging of the cosine squared. Therefore, in view of the earlier estimate in Eq.(S3-3),

$$\frac{|\overline{u_3}|}{|\widetilde{u_3}|} \approx 0.2. \tag{S3-4}$$

From the SAW structure it follows that all oscillating components are of the same order [6]: $\tilde{\eta}_3 \sim \tilde{\eta}_1$, while $\bar{\eta}_1 = 0$. The averaged components of the strain are generated by the averaged components of the thermo-elastic stress (which are proportional to the averaged (homogeneous) heating). Under this laterally homogeneous heating, there will be no lateral displacement of the surface, because the two directions are equivalent.

Therefore, the amplitude of the average component in our gRSAW is always much smaller than of the oscillating component. The estimate in (S3-4) is useful for a qualitative comparison of the different detection channels in the Supplemental material, Section 5, although a quantitative comparison requires the evaluation of the scattering integrals, which are different for the detection of the unmodulated component $\bar{\eta}_3$ and the oscillating components $\tilde{\eta}_3$ and $\tilde{\eta}_1$.

### 4. Modification of optical permittivity under the SAW strain field

Here we will consider how the components of the GaAs dielectric tensor $\varepsilon_{ij}$ are modified under the action of the SAW. To evaluate it we need to calculate the photo-elastic tensor $p_{ijkl}$ and the strain tensor $\eta_{ij}$, where $i, j$ = 1, 2, 3 denote the coordinate axes O$x_1$, O$x_2$, O$x_3$. The photo-elastic (opto-elastic) tensor is defined through the linear dependence of the components of the inversed dielectric tensor on the components of the strain tensor [7]: $\Delta(1/\varepsilon)_{ij} = p_{ijkl}\eta_{kl}$, which for our cubic crystal with dielectric constant $\varepsilon$ becomes: $\Delta\varepsilon_{ij} = -\varepsilon^2 p_{ijkl}\eta_{kl}$. In the contracted notations for cubic crystals, it can be written in the following form:

$$\begin{pmatrix} \Delta\varepsilon_1 \\ \Delta\varepsilon_2 \\ \\ \\ \\ \Delta\varepsilon_6 \end{pmatrix} \sim \begin{pmatrix} p_{11} & p_{13} & p_{15} \\ p_{12} & p_{23} & p_{25} \\ & & \\ & & \\ & & \\ p_{61} & p_{62} & p_{65} \end{pmatrix} \begin{pmatrix} \eta_1 \\ 0 \\ \eta_3 \\ 0 \\ \eta_5 \\ 0 \end{pmatrix}.$$

Here, we define $\Delta\varepsilon_1 \equiv \Delta\varepsilon_{11}$, $\Delta\varepsilon_2 \equiv \Delta\varepsilon_{yy}$ and $\Delta\varepsilon_6 \equiv \Delta\varepsilon_{12} = \Delta\varepsilon_{21}$ and consider only nonzero components of the strain in the SAW generated by the grating: $\eta_1 = \frac{\partial u_1}{\partial x_1}, \eta_3 = \frac{\partial u_3}{\partial x_3}, \eta_5 = \frac{\partial u_1}{\partial x_3} + \frac{\partial u_3}{\partial x_1}$.

Then we get:

$$\Delta\varepsilon_1 \sim (p_{11}\eta_1 + p_{13}\eta_3 + p_{15}\eta_5),$$
$$\Delta\varepsilon_2 \sim (p_{21}\eta_1 + p_{23}\eta_3 + p_{35}\eta_5),$$
$$\Delta\varepsilon_6 \sim (p_{61}\eta_1 + p_{63}\eta_3 + p_{65}\eta_5).$$

The photo-elastic tensor is known for the coordinate system related to the axes of the cubic crystal [8]. By transforming this tensor, $p'_{ij}$, we will evaluate the components of the photo-elastic tensor $p_{ij}$ in our experimental coordinate system, in which the axes $Ox_{1,2}$ are along the [110] and [−110] crystallographic axes of GaAs, while $Ox_3$ is along the [001] axis (see Fig. 1 in the main text):

$$p_{11} = p'_{11} - \left(\frac{p'_{11} - p'_{12}}{2} - p'_{44}\right),$$
$$p_{12} = p'_{12} + \left(\frac{p'_{11} - p'_{12}}{2} - p'_{44}\right),$$
$$p_{13} = p_{23} = p'_{12},$$
$$p_{15} = p_{25} = p_{61} = p_{63} = p_{65} = 0$$

Therefore, for change of optical permittivity under the action of the SAW we finally obtain:

$$\begin{aligned} \Delta\varepsilon_1 &= -\varepsilon^2\left\{\left[p'_{11} - \left(\frac{p'_{11} - p'_{12}}{2} - p'_{44}\right)\right]\eta_1 + p'_{12}\eta_3\right\}, \\ \Delta\varepsilon_2 &= -\varepsilon^2\left\{\left[p'_{12} + \left(\frac{p'_{11} - p'_{12}}{2} - p'_{44}\right)\right]\eta_1 + p'_{12}\eta_3\right\}, \\ \Delta\varepsilon_6 &= 0. \end{aligned} \quad \text{(S4-1)}$$

## 5. Optical detection of GHz-frequency SAWs

In picosecond ultrasonics [9] the detection of coherent acoustic waves inside the sample is based on heterodyning of probe laser pulses, scattered by the acoustic waves, by probe laser pulses, reflected at the stationary surface/interfaces of the sample. Both, acoustically scattered light transmitted from the sample into air and reflected light should be combined at a photodetector, where their electrical fields $\vec{E}^s \equiv \vec{E}_{t\leftarrow,s}$ and $\vec{E}_r$ interfere. The intensity of light on the photodetector reads

$$I \sim \vec{E}_r\left(\vec{E}_r\right)^* + 2Re\left[\left(\vec{E}_r\right)^*\vec{E}_{t\leftarrow,s}\right] + \vec{E}_{t\leftarrow,s}\left(\vec{E}_{t\leftarrow,s}\right)^*. \quad \text{(S5-1)}$$

As the acoustically scattered probe light intensity is much weaker than the reflected, the last term is negligibly small. The first term does not depend on the acoustic processes in the sample. It is the second term, which contains information on the acoustic waves. In case of a periodic modulation of the sample parameters the detection process acquires specific features, caused by the fact that the processes of reflection and scattering of the probe light are accompanied by its diffraction.

Although the problem can be solved in the general case, we will discuss here the situation, which corresponds to our experimental situation, where the sample parameters are periodic along a single direction (along the O$x_1$ axis, see Fig. 1 of the main text) due to the periodicity of the deposited metal grating and the incident probe light is directed along the $x_3$ axis, i.e., normal to the sample surface. The reflected probe light at the surface is described by $E_{r,i}(x_3 = 0) = (\bar{r}_{ij} + \tilde{r}_{ij})E_{0,j}$ , where $i,j = 1, 2$ denote the coordinates $x_{1,2}$, $E_{0,j}$ denotes the electric field components of the incident probe light, while the tensor of the electrical reflectivity is decomposed into its spatially homogeneous (averaged over the sample surface) part $\bar{r}_{ij}$ and the part $\tilde{r}_{ij}$, which is modulated along $x_1$ with the period of the metal grating, $a_{\text{gr}}$. The Fourier transform of the reflected light field along $x_1$ contains, in general, an infinite number of discrete components with wave numbers $k_{1,m} = \frac{2\pi}{d} m, m = 0, \pm1, \pm2, \ldots.$ along $x_1$. This means that the reflected light is decomposed in planes waves, directed in different diffraction orders. The propagation constant of light along the $x_3$ direction in the $m^{\text{th}}$ diffraction order is equal to $k_{3,m} = \sqrt{k_0^2 - k_{1,m}^2} = \sqrt{\left(\frac{2\pi}{\lambda_0}\right)^2 - \left(\frac{2\pi}{d} m\right)^2}$, where $k_0$ and $\lambda_0 \equiv \lambda_{\text{pr}}$ denote the wavevector and the wavelength of the probe light in air. Therefore, in our case of sub-optical gratings, i.e., $a_{\text{gr}} < \lambda_0$, only normally reflected light in the $m = 0$ order propagates, while the reflected light in the other diffraction orders is evanescent and cannot reach the detector, placed at a macroscopic distance from the sample. A similar situation takes place with the acoustically scattered probe light, which is transmitted into air. Therefore, at the photodetector only the reflected light and the light acoustically scattered and transmitted into air, which both propagate in the zeroth diffraction order, interfere.

The acoustically-scattered probe light, transmitted from the sample into air, can be described by $E_{t_{\leftarrow},s,i}(x_3 = 0) = (\bar{t}_{\leftarrow ik} + \tilde{t}_{\leftarrow ik})(\bar{s}_{kl} + \tilde{s}_{kl})(\bar{t}_{\rightarrow lj} + \tilde{t}_{\rightarrow lj})E_{0,j}$, where the periodically modulated tensor coefficients for light transmission from air into the substrate, $t_{\rightarrow lj}$, from the substrate into air, $t_{\leftarrow ik}$, and of the acoustically-induced scattering, $s_{kl}$, are introduced. Note that the scattering coefficients in this formula are evaluated at $x_3 = 0$ inside the substrate, i.e., before the transmission of the scattered light into air through the sample surface covered by the metal grating. They are integral results of the probe light scattering by acoustically-induced inhomogeneities of the optical permittivity in the complete volume of the sample, as will be mathematically formulated below. We will consider only the coefficients that can provide a nonzero contribution after averaging over the grating period, as they describe the probe light transmitted into air in the zeroth diffraction order, which reaches the photodetector:

$$E_{t_{\leftarrow},s,i}(x_3 = 0) = (\bar{t}_{\leftarrow ik}\bar{s}_{kl}\bar{t}_{\rightarrow lj} + \bar{t}_{\leftarrow ik}\overline{\tilde{s}_{kl}\tilde{t}_{\rightarrow lj}} + \bar{s}_{kl}\overline{\tilde{t}_{\leftarrow ik}\tilde{t}_{\rightarrow lj}} + \overline{\tilde{t}_{\leftarrow ik}\tilde{s}_{kl}}\bar{t}_{\rightarrow lj} + \overline{\tilde{t}_{\leftarrow ik}\tilde{s}_{kl}\tilde{t}_{\rightarrow lj}})E_{0,j}.$$

As follows from the above discussion the transmission/scattering coefficients with tildes "$\tilde{}$" and with bars "$\bar{}$" describe the processes with and without diffraction, respectively. Therefore, the derived presentation accounts for probe light, which never diffracts (first term inside the round brackets) or returns to the zeroth diffraction order after two or three diffraction processes (the terms 2–4 and 5, respectively). It is important to notice here, that although the probe light (transmitted into the sample and scattered by the acoustic field) is directed in different diffraction orders and can be either propagating or evanescent, its different contributions to scattering should be all carefully estimated, because the probed surface acoustic waves are localized at sub-micrometer distances from the surface.

However, we have the opportunity to simplify the above solution for the acoustically scattered probe light in our particular experimental situation of modulation introduced by the 10-nm-thick gold layer. We estimated that the amplitude of the modulated parts of the transmission coefficients are much smaller than their averaged parts, i.e., $|\tilde{t}| \leq 0.1|\bar{t}|$ (see Supplemental material, Section 6). Therefore, in the above solution $\left|\bar{s}_{kl}\overline{\tilde{t}_{\leftarrow\imath k}\tilde{t}_{\rightarrow l\jmath}}\right| \ll \left|\bar{t}_{\leftarrow ik}\bar{s}_{kl}\bar{t}_{\rightarrow lj}\right|$, $\left|\overline{\tilde{t}_{\leftarrow\imath k}\tilde{s}_{kl}\tilde{t}_{\rightarrow l\jmath}}\right| \ll \left|\bar{t}_{\leftarrow ik}\overline{\tilde{s}_{kl}\tilde{t}_{\rightarrow l\jmath}} + \overline{\tilde{t}_{\leftarrow\imath k}\tilde{s}_{kl}}\bar{t}_{\rightarrow lj}\right|$ and it takes the form:

$$E_{t_{\leftarrow},s,i}(x_3 = 0) = (\bar{t}_{\leftarrow ik}\bar{s}_{kl}\bar{t}_{\rightarrow lj} + \bar{t}_{\leftarrow ik}\overline{\tilde{s}_{kl}\tilde{t}_{\rightarrow l\jmath}} + \overline{\tilde{t}_{\leftarrow\imath k}\tilde{s}_{kl}}\bar{t}_{\rightarrow lj})E_{0,j}. \tag{S5-2}$$

This solution can be simplified additionally for light scattered by the bulk quasi-longitudinal acoustic (QLA) waves emitted by the grating. The weak modulation of the light transmission coefficients, estimated in the Supplemental material, Section 6, is intrinsically related in our sample to the weak modulation of the reflection coefficients and of the pump light absorption. In accordance with [4], the gold layer of 10 nm thickness absorbs only 10% of light at 800 nm wavelength, confirming our estimates. Due to the weak modulation of the pump light absorption by the optically thin metallic grating, the QLA waves, launched in the non-zeroth diffraction orders by the thermoelastic mechanism of optoacoustic conversion, are much weaker than one emitted normally to the sample surface. Therefore, in the probe light scattered by the LA waves the scattering by the purely LA plane wave, emitted by the metal grating in the zeroth diffraction order, i.e., along $x_3$ axis largely dominates:

$$E^{LA}_{t_{\leftarrow},s,i}(x_3 = 0) = \bar{t}_{\leftarrow ik}\bar{s}_{kl}\bar{t}_{\rightarrow lj}E_{0,j}. \tag{S5-3}$$

This conclusion is confirmed by our experimental observations of time-domain Brillouin scattering (TDBS) [9, 10] at the frequency of ~40 GHz suggested by the detection process in Eq. (S5-3).

For the detection of the SAWs the simplification of Eq. (S5-2), based on an inequality similar to $\left|\tilde{s}_{QLA}\right| \ll \left|\bar{s}_{QLA}\right|$, is not valid. In this case the average and oscillating components of the scattering coefficients are controlled by the spatially averaged (laterally homogeneous) and laterally oscillating components of the generalized Rayleigh SAW (gRSAW, a surface-localized eigen mode of the laterally modulated elastic structures). Our experimental observations suggest a weak dispersion and a small diminishing of the gRSAW velocity caused by loading of the sample surface by the metallic grating (see Fig. 2(c) in the main text). Even for the shortest grating period the velocity diminishes only by ~17%. The estimates for our experimental case of weak modulation due to the thin gold stripes (see Supplemental material, Section 6 and [6]) suggest, that $\left|\bar{s}_{gRSAW}\right| \leq 0.1\left|\tilde{s}_{gRSAW}\right|$. Therefore, potentially, all three detection paths, suggested by Eq. (S5-2), should be carefully considered.

Figure 5 of the main text depicts three schematic paths for the SAW detection. The oscillations of the homogeneous part of the electrical permittivity, caused by the homogeneous part of the strain field in gRSAW are detectable by probe light propagating in the zeroth diffraction order (Fig. 5(c)), while the spatially modulated components of the electrical permittivity are detectable due to the probe light diffraction processes (Fig. 5(a),(b)). Either the plane probe light is first scattered in the non-zeroth diffraction orders by the acoustical field and then diffracted to the direction of the photodetector (heterodyning direction) by the metallic grating in transmission from the sample to air (third term in Eq. (S5-2)), or the opposite, i.e., probe light is first diffracted in transmission from air into the sample and then returned to the heterodyning direction via scattering by the surface acoustic wave (second term in Eq. (S5-2)).

Additional significant simplifications of the above derived solution are valid for our experiments. To start, we account for those, which are fruitful for the understanding of the detection of both the longitudinal and surface waves. First, due to the grating structure and optical isotropy of the sample materials, the tensors of the reflection and the transmission coefficient are diagonal. Moreover, due to the chosen orientation of the grating and, thus, the direction of the gRSAW propagation, along the [110] GaAs crystallographic axis, the tensor of the acoustically-induced polarization of the material and, consequently, the scattering tensors in Eq. (S5-2), are also diagonal (see the Supplemental material, Section 4). Therefore, the components $E_{0,1}$ and $E_{0,2}$ of the probe light incident on the sample interact with the sample independently and are not coupled. So it is possible to replace the tensors with scalars, although different in magnitude for the two different orientations of the electric field. Considering that in our experimental scheme, the electric field of the incident probe light is oriented at $\alpha = 45°$ to the lateral coordinate axes: $E_{0,i} \equiv \frac{1}{\sqrt{2}} E_0$. Then the components of the reflected and acoustically scattered electric field transmitted into air, both directed to the detector, can be written as follows:

$$E_{r,i}(x_3 = 0) = \frac{1}{\sqrt{2}} \bar{r}_i E_0,\ E_{t_{\leftarrow},s,i}^{LA}(x_3 = 0) = \frac{1}{\sqrt{2}} \bar{t}_{\leftarrow i} \bar{s}_i \bar{t}_{\rightarrow i} E_0, \tag{S5-4}$$

$$E_{t_{\leftarrow},s,i}^{gRSAW}(x_3 = 0) = \frac{1}{\sqrt{2}} \left(\bar{t}_{\leftarrow i} \bar{s}_i \bar{t}_{\rightarrow i} + \bar{t}_{\leftarrow i} \overline{\tilde{s}_\iota \tilde{t}_{\rightarrow \iota}} + \overline{\tilde{t}_{\leftarrow \iota} \tilde{s}_\iota} \bar{t}_{\rightarrow i}\right) E_0,\ \text{where } i = 1,2.$$

In our detection scheme we measure the polarization rotation of the acoustically-scattered probe light transmitted from the sample to the detector. For this, the difference in intensities of the two orthogonal electric field components is measured by the balanced photodetector. Taking into account that two orthogonal components of the electric field of the probe are interacting with the sample independently, we obtain for their intensities:

$$I_i \cong E_{r,i} \left(E_{r,i}\right)^* + 2Re\left[\left(E_{r,i}\right)^* E_{s,i}\right], \tag{S5-5}$$

where $E_{s,i}$ denotes the sum of the electric field components scattered by the LA wave and gRSAW presented in Eq. (S5-4). The scheme is balanced in the absence of the acoustic waves and acoustically-scattered probe light, i.e., when blocking the pump laser pulses. This is fulfilled by rotation of the polarization of the probe light, transmitted in the direction of the photodetector, by an angle $\delta\alpha$. The components of the electric field after the rotation become $E_{r,s,1}^{\delta\alpha} = E_{r,s,1} \cos(\delta\alpha) - E_{r,s,2} \sin(\delta\alpha)$, $E_{r,s,2}^{\delta\alpha} = E_{r,s,1} \sin(\delta\alpha) + E_{r,s,2} \cos(\delta\alpha)$, while their intensities in (S5-5) can be rewritten as

$$I_i^{\delta\alpha} \cong E_{r,i}^{\delta\alpha} \left(E_{r,i}^{\delta\alpha}\right)^* + 2Re\left[\left(E_{r,i}^{\delta\alpha}\right)^* E_{s,i}^{\delta\alpha}\right]. \tag{S5-6}$$

The polarization rotation angle of the acoustically-scattered probe light $\vartheta_R$ is proportional to the normalized differential signal measured by the balanced photodetector:

$$\vartheta_R \sim S = \frac{I_1^{\delta\alpha} - I_2^{\delta\alpha}}{I_1^{\delta\alpha} + I_2^{\delta\alpha}} = \frac{\left[E_{r,1}^{\delta\alpha}\left(E_{r,1}^{\delta\alpha}\right)^* - E_{r,2}^{\delta\alpha}\left(E_{r,2}^{\delta\alpha}\right)^*\right] + 2\mathrm{Re}\left[\left(E_{r,1}^{\delta\alpha}\right)^* E_{s,1}^{\delta\alpha} - \left(E_{r,2}^{\delta\alpha}\right)^* E_{s,2}^{\delta\alpha}\right]}{\left[E_{r,1}^{\delta\alpha}\left(E_{r,1}^{\delta\alpha}\right)^* + E_{r,2}^{\delta\alpha}\left(E_{r,2}^{\delta\alpha}\right)^*\right] + 2\mathrm{Re}\left[\left(E_{r,1}^{\delta\alpha}\right)^* E_{s,1}^{\delta\alpha} + \left(E_{r,2}^{\delta\alpha}\right)^* E_{s,2}^{\delta\alpha}\right]}. \tag{S5-7}$$

The scheme is balanced in the absence of the acoustic waves and acoustically-scattered probe light, i.e., when blocking the pump laser pulses, leading to $E_{s,1}^{\delta\alpha} = E_{s,2}^{\delta\alpha} = 0$ in Eq. (S5-7). Thus, the angle of rotation $\alpha$ is determined by the condition $E_{r,1}^{\delta\alpha}\left(E_{r,1}^{\delta\alpha}\right)^* - E_{r,2}^{\delta\alpha}\left(E_{r,2}^{\delta\alpha}\right)^* = 0$, which gives

$$\left(\left|E_{r,1}\right|^2 - \left|E_{r,2}\right|^2\right) \cos(2\delta\alpha) - \left(E_{r,1} E_{r,2}^* + E_{r,1}^* E_{r,2}\right) \sin(2\delta\alpha) = 0. \tag{S5-8}$$

It is instructive to estimate the required tuning angle $\delta\alpha$ in our experiments. Introducing the modulus and the phase of the reflection coefficient, $\bar{r}_i \equiv |\bar{r}_i| e^{i\varphi_{r,i}}$, the formula for the rotation angle takes the form:

$$\tan(2\delta\alpha) = \frac{\left|E_{r,1}\right|^2 - \left|E_{r,2}\right|^2}{E_{r,1}E_{r,2}^* + E_{r,1}^*E_{r,2}} = \frac{|\bar{r}_1|^2 - |\bar{r}_2|^2}{2|\bar{r}_1||\bar{r}_2|\cos(\varphi_{r,1} - \varphi_{r,2})} = \frac{\left|\frac{I_1}{I_2}\right|^2 - 1}{2\left|\frac{I_1}{I_2}\right|\cos(\varphi_{r,1} - \varphi_{r,2})} \cong \left|\frac{I_1}{I_2}\right| - 1 \,.$$

The final approximation above comes from the expectations and the estimates, suggesting that thin gold stripes weakly perturb the reflectivity of the probe light (both its amplitude and phase). These expectations are supported by the results of the measurements, which demonstrate that the anisotropy of the signal is relatively small: $\left|\frac{I_1}{I_2}\right| - 1 \leq 0.25$. Therefore, the rotation angle is also small, i.e., $\delta\alpha \leq 7^{\circ}$, $\cos(2\delta\alpha) \geq 0.97$.

When the scheme is balanced, the signal is proportional to a small term, retained in the nominator of Eq. (S5-7), while the second small contribution "from acoustics" in the denominator of Eq. (S5-7) can be neglected:

$$S = \frac{2\mathrm{Re}\left[\left(E_{r,1}^{\delta\alpha}\right)^* E_{s,1}^{\delta\alpha} - \left(E_{r,2}^{\delta\alpha}\right)^* E_{s,2}^{\delta\alpha}\right]}{\left[E_{r,1}^{\delta\alpha}\left(E_{r,1}^{\delta\alpha}\right)^* + E_{r,2}^{\delta\alpha}\left(E_{r,2}^{\delta\alpha}\right)^*\right]} \cong \frac{2\mathrm{Re}\left[\left(E_{r,1}^{\delta\alpha}\right)^* E_{s,1}^{\delta\alpha} - \left(E_{r,2}^{\delta\alpha}\right)^* E_{s,2}^{\delta\alpha}\right]}{r_0^2\, E_0^2}.$$

Here, for compactness of the following formulas, we have neglected in the denominator all small corrections, induced by small deviations of the reflection coefficients from the real reflection coefficient $r_0$ of the bare optically isotropic GaAs substrate. However, in principle, the denominator is just the intensity of the reflected light, i.e., a real independent of the acoustics coefficient, which does not influence the nature and relative importance of the physical processes described by the nominator. Explicitly using Eq. (S5-8), the signal can be subsequently evaluated as

$$S = \frac{2\cos(2\delta\alpha)\,\mathrm{Re}\left\{\left[\left(E_{r,1}^*\right)^2 + \left(E_{r,2}^*\right)^2\right]\left(E_{r,2}E_{s,1} - E_{r,1}E_{s,2}\right)\right\}}{r_0^2 E_0^2\left(E_{r,1}E_{r,2}^* + E_{r,1}^*E_{r,2}\right)} \cong \frac{2\mathrm{Re}\left(E_{r,2}E_{s,1} - E_{r,1}E_{s,2}\right)}{r_0^2\, E_0^2}, \tag{S5-9}$$

Before applying Eqs. (S5-9) to our particular acoustic fields, it is insightful to consider it in a hypothetical situation where the metal grating is absent. In this case, the transmission and reflection coefficients for the components $E_{0,1}$ and $E_{0,2}$ of the probe light are equal, predicting that our experimental scheme is sensitive only to the acoustic field components, which provide an anisotropic photo-elastic response, i.e., by differently scattering the components $E_{0,1}$ and $E_{0,2}$, and are capable to rotate the probe light electric field vector. Therefore, our derived formula Eq. (S5-9) suggests that the acoustic field components, which are not rotating the probe electric field can be potentially detected in our polarization-sensitive scheme only due to the optically anisotropy induced by the metal grating.

We now apply the derived compact formula (S5-9) to the evaluation of the signal resulting from the scattering by the SAW, which is presented in Eq. (S5-4). The first term for the probe light, scattered by the SAW, is the same as for the probe light scattered by the LA wave. Therefore, the case of the LA wave, is included in our analysis.

$$S = \frac{2\mathrm{Re}\{[E_{r,2}E_{s,1} - (2\leftrightarrow 1)]\}}{r_0^2\, E_0^2} = \frac{1}{r_0^2}\mathrm{Re}\left[\bar{r}_2\left(\bar{t}_{\leftarrow 1}\bar{s}_1\bar{t}_{\rightarrow 1} + \bar{t}_{\leftarrow 1}\overline{\tilde{s}_1\tilde{t}_{\rightarrow 1}} + \overline{\tilde{t}_{\leftarrow 1}\tilde{s}_1}\bar{t}_{\rightarrow 1}\right) - (2\leftrightarrow 1)\right]. \tag{S5-10}$$

For further compact evaluation of the different contributions in Eq. (S5-10), in the description of the spatial inhomogeneity of the transmission coefficients and of the acoustic field we will consider only the first harmonics of their spatial spectra. From symmetry of the gratings with a duty cycle of 50%, it follows that the strain components $\eta_1$ and $\eta_3$, relevant for our geometry (see Supplemental material, Section 3) can be both presented as $\eta_l = \overline{\eta_l}(x_3) + \widetilde{\eta_l}(x_3)\cos\left(k_g x_1\right) \equiv \eta_{l,m}\cos\left(m k_g x_1\right)$, where $l = 1, 3$ denotes the coordinate and $m = 0,1$ is the summation index. Then the SAW-induced change of the optical permittivity is $\Delta\varepsilon_l = -\varepsilon^2 p_{lj}\eta_j = \overline{\Delta\varepsilon_l}(x_3) + \widetilde{\Delta\varepsilon_l}(x_3)\cos\left(k_g x_1\right) \equiv \Delta\varepsilon_{l,m}(x_3)\cos\left(m k_g x_1\right)$. The grating-induced variations of the transmission coefficients in Eq.(S5-10) can also be presented as $t_l = \bar{t}_{l,m} + \tilde{t}_{l,m}\cos(k_g x_1) \equiv t_{l,m}\cos(mk_g x_1)$, where $k_g$ denotes the grating wave number.

Note, that the amplitude of the next term in the Fourier expansion of the rectangular spatial distribution with 50% duty cycle is 3 times smaller than that of the first harmonic. The intensity of the probe light, which exhibits double diffraction (to be directed to the detector) is proportional to the square of this component, and therefore is an order of magnitude weaker than the one evaluated below. Finally, a strong argument for neglecting the third spatial harmonic is that the probe light directed in the corresponding diffraction order is evanescent and mostly does not reach the GaAs excitonic layer. We estimated that the attenuation depth of probe light in the $\pm 3$ diffraction orders is $\leq 15$ nm for a grating period $a_{\mathrm{gr}} \leq 340$ nm.

We present now the derivation of the three coefficients of different type, which contribute to the signal in Eq. (S5-10), i.e., $\bar{t}_{\leftarrow 1}\bar{s}_1\bar{t}_{\rightarrow 1}$, $\bar{t}_{\leftarrow 1}\overline{\tilde{s}_1\tilde{t}_{\rightarrow 1}}$, and $\overline{\tilde{t}_{\leftarrow 1}\tilde{s}_1}\bar{t}_{\rightarrow 1}$. The reflections of the probe light at the interfaces of GaAs and AlGaAs are negligibly small in our sample: $|r| \cong 0.03$ (the refractive index of $Al_{0.34}Ga_{0.66}As$ at 820 nm is ~3.4 [11]). Therefore, assuming our sample to be optically homogenous, the propagation of probe light is perturbed only by the presence of the acoustic field described by the equation

$$\left(\frac{\partial^2}{\partial x_1^2} + \frac{\partial^2}{\partial x_3^2}\right) E_1 + k_0^2 \varepsilon \left[1 + \frac{\Delta\varepsilon_1(x_1,x_3)}{\varepsilon}\right] E_1 = 0, \text{ or}$$

$$\left(\frac{\partial^2}{\partial x_1^2} + \frac{\partial^2}{\partial x_3^2}\right) E_1 + k_n^2 E_1 = -k_n^2 E_{a-i}, \qquad \text{(S5-11)}$$

where $k_n \equiv k_0\sqrt{\varepsilon} = 2\pi n/\lambda_0$ is the probe wave vector in the media (GaAs), $n = \sqrt{\varepsilon}$ is complex refractive index of GaAs at the probe wavelength $\lambda_0$. The small term $E_{a-i} \equiv \frac{\Delta\varepsilon_1(x_1,x_3)}{\varepsilon} E_1$ (on the right-hand-side) is called the acoustically-induced (*a-i*) nonlinear optical polarization. We remind that $E_1$ stands for the $x_1$ component of the probe light. Because of the smallness of the *a-i* polarization, Eq. (S5-11) can be solved in the single scattering approximation. The probe field launched in the sample ($x_3 \geq 0$) in the absence of acoustic waves can be presented as

$$E_{1\rightarrow} = t_{\rightarrow 1,m}\cos(mk_g x_1)\, e^{i\sqrt{k_n^2 - (mk_g)^2}\, x_3} E_{1,0}. \qquad \text{(S5-12)}$$

Here "0" indicates the electric field component of the incident probe light. This description, where $m$ =0, 1 is the summation index, includes both the probe light propagating in the first diffraction orders, i.e., $\pm 1$, when $m$ = 1 (spatially modulated term), and the non-diffracted light, when $m$ = 0 (non-modulated term). The solution (S5-12) satisfies Eq. (S5-11) in the absence of the acoustic field. Substituting it in the right-hand side of Eq. (S5-11), we derive the acoustically-induced polarization:

$$E_{a-i}(x_1,x_3) \equiv t_{\rightarrow 1,m}\frac{\Delta\varepsilon_1(x_1,x_3)}{\varepsilon}\cos(mk_g x_1)\, e^{i\sqrt{k_n^2-(mk_g)^2}\, x_3} E_{1,0} =$$

$$= t_{\rightarrow 1,m}\frac{1}{\varepsilon}\Delta\varepsilon_{1,m'}(x_3)\cos(m'k_g x_1)\cos(mk_g x_1)\, e^{i\sqrt{k_n^2-(mk_g)^2}\, x_3} E_{1,0}, \qquad \text{(S5-13)}$$

where we have introduced the summation index $m' = 0{,}1$ for the descriptionof the permittivity modulation. The $x_3$ distribution of $E_{a-i}(x_1,x_3)$ is different for the non-diffracted and diffracted probe light.

The acoustically-scattered probe light is the solution of (S5-11) with the $E_{a-i}$ presented in (S5-13). It can be given in the form of counter-propagating light waves of variable amplitudes:

$$E_{1\leftarrow,m''}(x_3)\cos\left(m''k_g x_1\right)e^{-i\sqrt{k_n^2-\left(m''k_g\right)^2}x_3}+E_{1\rightarrow,m''}(x_3)\cos\left(m''k_g x_1\right)e^{i\sqrt{k_n^2-\left(m''k_g\right)^2}x_3}.$$

Here $m''=0{,}1$ is the summation index. The amplitude of the light, scattered in the direction of the surface, at the surface ($x_3=0$), is given by the overlap integral

$$E_{1\leftarrow,m''}(0)=\frac{ik_n^2}{2\sqrt{k_n^2-\left(m''k_g\right)^2}}\int_0^\infty E_{a-i}(x_1,x_3)\,e^{i\sqrt{k_n^2-\left(m''k_g\right)^2}x_3}dx_3E_{1,0}=$$

$$t_{\rightarrow1,m}\frac{ik_n^2}{2\sqrt{k_n^2-\left(m''k_g\right)^2}}\int_0^\infty\frac{\Delta\varepsilon_{1,m'}(x_3)}{\varepsilon}\cos\left(m'k_g x_1\right)\cos\left(mk_g x_1\right)e^{i\left(\sqrt{k_n^2-\left(mk_g\right)^2}+\sqrt{k_n^2-\left(m''k_g\right)^2}\right)x_3}dx_3E_{1,0},\quad\text{(S5-14)}$$

which for a particular scattered wave in Eq. (S5-14) requires a choice in the $E_{a-i}(x_1,x_3)$ part, which emits this particular wave.

For the evaluation of the coefficient $\bar{t}_{\leftarrow1}\bar{s}_1\bar{t}_{\rightarrow1}$ in Eq. (S5-10), describing the light transmitted to the photodetector after scattering of the non-diffracted light wave into the non-diffracted light wave, the coefficient $\bar{s}_1\bar{t}_{\rightarrow1}$ provided by Eq. (S5-14) with $m=m'=m''=0$, should be additionally multiplied by $t_{\leftarrow1,0}$ :

$$\bar{t}_{\leftarrow1}\bar{s}_1\bar{t}_{\rightarrow1}=t_{\leftarrow1,0}t_{\rightarrow1,0}\frac{ik_n}{2}\int_0^\infty\frac{\Delta\varepsilon_{1,0}(x_3)}{\varepsilon}e^{2ik_nx_3}dx_3\equiv t_{\leftarrow1,0}t_{\rightarrow1,0}\frac{ik_n}{2}J_{p\rightarrow p,1}.\quad\text{(S5-15)}$$

In Eq. (S5-15), the overlap integral $J_{p\rightarrow p,1}\equiv J_{p,1}$ of the probe light with the depth distribution of the *a-i* permittivity changes takes a form, which is classic for picosecond laser ultrasonics with plane waves [9, 10]. For the $\overline{\tilde{s}\tilde{t}_{\rightarrow1}}$ scattering, the incident probe light in the first diffraction orders corresponds to $m=1$, the scattered light in the zeroth order corresponds to $m''=0$, while the respective amplitude of the *a-i* polarization is due to the lateral averaging of the cosines product with $m=m'=1$ in Eq. (S5-14). Multiplying this result by $t_{\leftarrow1,0}$, leads to

$$\bar{t}_{\leftarrow1}\overline{\tilde{s}_1\tilde{t}_{\rightarrow1}}=t_{\leftarrow1,0}t_{\rightarrow1,1}\frac{ik_n}{4}\int_0^\infty\frac{\Delta\varepsilon_{1,1}(x_3)}{\varepsilon}e^{i\left(\sqrt{k_n^2-k_g^2}+k_n\right)x_3}dx_3\equiv t_{\leftarrow1,0}t_{\rightarrow1,1}\frac{ik_n}{4}J_{d\rightarrow p,1}.\quad\text{(S5-16)}$$

Finally, the $\tilde{s}_1\bar{t}_{\rightarrow1}$ process involves diffraction by the grating of the non-perturbed probe light, i.e., $m=0, m'=m''=1$. The corresponding solution of Eq. (S5-14)

$$E_{1\leftarrow,1}(0)=t_{\rightarrow1,0}\frac{ik_n^2}{2\sqrt{k_n^2-k_g^2}}\cos\left(k_gx_1\right)\int_0^\infty\frac{\Delta\varepsilon_{1,1}(x_3)}{\varepsilon}e^{i\left(k_n+\sqrt{k_n^2-k_g^2}\right)x_3}dx_3E_{1,0},$$

should be multiplied by $t_{\leftarrow1,1}\cos\left(k_gx_1\right)$ and averaged over the lateral coordinate, resulting in:

$$\overline{\tilde{t}_{\leftarrow1}\tilde{s}_1}\bar{t}_{\rightarrow1}=t_{\leftarrow1,1}t_{\rightarrow1,0}\frac{ik_n^2}{4\sqrt{k_n^2-k_g^2}}\int_0^\infty\frac{\Delta\varepsilon_{1,1}(x_3)}{\varepsilon}e^{i\left(k_n+\sqrt{k_n^2-k_g^2}\right)x_3}dx_3=t_{\leftarrow1,1}t_{\rightarrow1,0}\frac{ik_n^2}{4\sqrt{k_n^2-k_g^2}}J_{p\rightarrow d,1}.\quad\text{(S5-17)}$$

The solutions in Eqs. (S5-16) and (S5-17) demonstrate that the two analyzed detection processes with participation of the diffracted light are described by the same overlap integral $J_{d\rightarrow p,1}=J_{p\rightarrow d,1}\equiv J_{d,1}$ of the probe light with a depth distribution of the *a-i* permittivity changes, which is different from the one in Eq. (S5-15).

The coefficients evaluated in Eqs.(S5-15) − (S5-17) are substituted in the theoretical formula for the signal in Eq. (S5-10):

$$S = \frac{1}{r_0^2}\,\mathrm{Re}\left\{\frac{ik_n}{2}\left[\bar{r}_2 t_{\leftarrow 1,0} t_{\rightarrow 1,0}\left(J_{p,1} + \frac{t_{\rightarrow 1,1}}{t_{\rightarrow 1,0}}\frac{J_{d,1}}{2} + \frac{t_{\leftarrow 1,1}}{t_{\leftarrow 1,0}}\frac{J_{d,1}}{2\sqrt{1-\left(k_g/k_n\right)^2}}\right)\right] - (2 \leftrightarrow 1)\right\}. \quad \text{(S5-18)}$$

From classical relations between the transmission and reflection coefficients for the interfaces and embedded layers, it follows $t_{\leftarrow 1,0} t_{\rightarrow 1,0} = 1 - \bar{r}_1^2$ (where $\bar{r}_1$ denotes the laterally averaged reflection coefficient of the $x_1$ component of the electric field from the sample surface), while the modulation coefficients of the transmission coefficients don't depend on the direction of light propagation, i.e., $\frac{t_{\rightarrow 1,1}}{t_{\rightarrow 1,0}} = \frac{t_{\leftarrow 1,1}}{t_{\leftarrow 1,0}} \cong \frac{\tilde{t}_1}{\bar{t}_1} \equiv m_1$.

Eq. (S5-18) takes the form

$$S = -\frac{k_n}{2r_0^2}\,\mathrm{Im}\left\{\bar{r}_2\left(1-\bar{r}_1^2\right)\left[J_{p,1} + c m_1 J_{d,1}\right] - (2 \leftrightarrow 1)\right\}, \quad \text{(S5-19)}$$

where the compact notation $c \equiv \frac{1}{2}\left(1 + \frac{1}{\sqrt{1-\left(k_g/k_n\right)^2}}\right)$ is introduced for the coefficient, which is independent on the orientation of the probe electric field. It is important to note, that when the probe wavelength in the medium is comparable with the grating period, and thus $k_g \approx k_n$, significant influence on the signal amplitude has the imaginary part of the complex refractive index $n$, i.e., absorption of the probe light.

It is now instructive to split all the functions and parameters, $f$, contributing to the above signal, into their parts which are isotropic, $f_+ = \frac{f_2+f_1}{2}$, i.e., are the same for both orientations of the electric field, and anisotropic, $f_- = \frac{f_2-f_1}{2}$, i.e., differ by a sign for the two orientations of the electric field: $f_{1,2} = f_+ \mp f_-$.

$$S \cong \frac{k_n}{r_0}\left(1-r_0^2\right)\mathrm{Im}\left\{\left[J_{p-} + c\left(m_+ J_{d-} + m_- J_{d+}\right)\right] - \frac{\left(1+r_0^2\right)}{\left(1-r_0^2\right)}\left(\frac{r_-}{r_0}\right)\left[J_{p+} + c\left(m_+ J_{d+} + m_- J_{d-}\right)\right]\right\}. \quad \text{(S5-20)}$$

In Eq. (S5-20), as sometimes earlier, we omitted all negligible corrections to the coefficients $r_0\left(1-r_0^2\right)$ and $\left(1+r_0^2\right)$, which appear due to the small deviations of the reflection coefficient from that of bare GaAs ($r_0$).

The optical permittivity of the sample is modified by the acoustic strain of the SAWs in the following way (see Eq. (S4-1) of the Supplemental material, Section 4):

$$\Delta\varepsilon_1 = -\varepsilon^2\left\{\left[p'_{11} - \left(\frac{p'_{11}-p'_{12}}{2} - p'_{44}\right)\right]\eta_1 + p'_{12}\eta_3\right\},$$

$$\Delta\varepsilon_2 = -\varepsilon^2\left\{\left[p'_{12} + \left(\frac{p'_{11}-p'_{12}}{2} - p'_{44}\right)\right]\eta_1 + p'_{12}\eta_3\right\},$$

Therefore, $J_+ \sim \Delta\varepsilon_+ \sim \left[\left(\frac{p'_{11}+p'_{12}}{2}\right)\right]\eta_1 + p'_{12}\eta_3$, while $J_- \sim \Delta\varepsilon_- \sim -p'_{44}\eta_1$.

The solution (S5-20) suggests that the isotropic part of optical reflectivity allows detecting acoustic waves, which are rotating the electrical field vector, without assistance of the diffraction grating. The corresponding contribution to the signal in (S5-20) is:

$$S_{p-} = \frac{k_n}{r_0}\left(1-r_0^2\right)Im\left\{J_{p-}\right\}. \quad \text{(S5-21)}$$

However, from the symmetry of the grating with 50% duty cycle it follows, that only $\eta_3$ can contain a laterally homogeneous (nonzero averaged) part, detectable without diffracted probe light, while $\bar{\eta}_1 = 0$. So we immediately get:

$$J_{p+} \sim p'_{12}\bar{\eta}_3 \neq 0,$$

$$J_{p_} \sim -p'_{44}\bar{\eta}_1 = 0.$$

Therefore, the detection path in Eq. (S5-21) is inactive and the signal detection in our experimental configuration is exclusively due to the presence of the grating:

$$S \cong \frac{k_n}{r_0}(1-r_0^2)\mathrm{Im}\left\{[c(m_+J_{d-}+m_-J_{d+})] - \frac{(1+r_0^2)}{(1-r_0^2)}\left(\frac{r_-}{r_0}\right)[J_{p+}+c(m_+J_{d+}+m_-J_{d-})]\right\}. \quad \text{(S5-22)}$$

Eq. (S5-21) predicts, that if the optical anisotropy of the grating (caused by the difference in the finite width of the metal stripe along the $x_1$ direction and its infinite length in the $x_2$ direction and dominantly due to the plasmon resonance effect, possible for the $x_1$ component of the electric field) is neglected, then, similar to (S5-21), only the acoustic waves rotating the probe electric field are detectable even with the assistance of the metal grating:

$$S_{d-} = \frac{k_n}{r_0}(1-r_0^2)\mathrm{Im}\{cm_+J_{d-}\}. \quad \text{(S5-23)}$$

*In the absence of the grating-induced optical anisotropy, all the other detection paths* in Eq. (S5-22) *are inactive* because either $r_- = 0$ or $m_- = 0$, or both.

Optical anisotropy of the grating, i.e., $r_- \neq 0$ or $m_- \neq 0$, in accordance with Eq. (S5-22) provides the opportunity to detect both acoustic waves, which are rotating (terms $\sim J_{d-}$) and not rotating (terms $\sim J_{p+}$, $\sim J_{d+}$) the electric field vector of the probe light. However, our estimates in the Supplemental material, Section 6 indicate that $|m_-| \ll |m_+|$ and, additionally, the anisotropy of the optical reflectivity caused by the plasmon effect is weak, $\left|\left(\frac{r_-}{r_0}\right)\right| \leq 0.05$. Therefore, the scattering process $\sim J_{d-}$ is dominantly detected via the path in Eq. (S5-23). Our formula for the detection process becomes:

$$S \cong \frac{k_n}{r_0}(1-r_0^2)\mathrm{Im}\left\{c\left[m_+J_{d-}+\left(m_- - \frac{(1+r_0^2)}{(1-r_0^2)}\left(\frac{r_-}{r_0}\right)m_+\right)J_{d+}\right] - \frac{(1+r_0^2)}{(1-r_0^2)}\left(\frac{r_-}{r_0}\right)J_{p+}\right\}, \quad \text{(S5-24)}$$

where the two paths to access the scattering process $\sim J_{d+}$ are estimated to be competitive. We remind here, that $J_{d-}\sim\Delta\tilde{\varepsilon}_- = (-p'_{44}\tilde{\eta}_1)$, $J_{d,+}\sim\Delta\tilde{\varepsilon}_+\sim\left[\left(\frac{p'_{11}+p'_{12}}{2}\right)\right]\tilde{\eta}_1 + p'_{12}\tilde{\eta}_3$, while $J_{p+}\sim p'_{12}\tilde{\eta}_3$. Further comparison of the different contributions to the detection of the SAWs requires taking into account its structure and information on the photo-elastic constants. From the structure of the SAWs [5] it follows that the components of strain $\tilde{\eta}_1$ and $\tilde{\eta}_3$ have the opposite sign, while their depth distributions are rather similar. At the same time the information on the photo-elastic constants of GaAs in its transparency region [12] suggests that the different photo-elastic constants are of the same order and sign. Therefore, presumably, $J_{d-}$ and $J_{d+}$ are comparable, while the coefficients in front of them are strongly different, leading to the following simplification of the detected signal formula in Eq. (S5-24):

$$S \equiv (S_d + S_p) \cong \frac{k_n}{r_0}(1-r_0^2)\mathrm{Im}\left\{cm_+J_{d-} - \frac{(1+r_0^2)}{(1-r_0^2)}\left(\frac{r_-}{r_0}\right)J_{p+}\right\}. \quad \text{(S5-25)}$$

Note that for convenience, in the main text the isotropic part of the relative transmission coefficient $m_+ \equiv \frac{1}{2}\left(\frac{\tilde{t}_1}{\bar{t}_1}+\frac{\tilde{t}_2}{\bar{t}_2}\right)$ is denoted by $\tilde{t}_+$, while $\bar{r}_-$ denotes the anisotropic part of the reflection tensor: $\bar{r}_- \equiv \frac{r_-}{r_0}$.

In Eq. (S5-25) it is tempting to neglect the second term, because in our gRSAW the laterally homogeneous component of the strain, probed by $J_{p+}$, is much smaller than the modulated component of the strain, probed by $J_{d-}$ (see Supplemental material, Section 3). However, it is impossible to do this straightforward, because the detection of this component does not require the diffracted probe light, whereas the detection of the laterally modulated component does, while the diffracted probe light can be evanescent, failing to reach the exciton layer with the enhanced photo-elastic response.

In fact, our diffracted probe light with the wavelength of about $\lambda_n = \frac{\lambda_0}{n} \approx 225$ nm in GaAs is propagating in the $x_3$ direction in half of our gratings (with the periods 260, 340 and 400 nm) and is evanescent in the $x_3$ direction for the other half of gratings, i.e., with the periods 100, 160 and 220 nm, where it penetrates to the characteristic depths $l_p \equiv \frac{\lambda_n}{2\pi\sqrt{\left(\frac{\lambda_n}{d}\right)^2 - 1}} \approx 18, 24, \text{and } 167$ nm, respectively. Therefore, a strong diminishing of the signals described by the first contribution to Eq. (S5-25) is expected in the two shortest gratings. The diffracted probe light practically does not reach the exciton layer (located at depths from 15 nm to 65 nm) in the case of the 100 nm grating and probes just a small part of it in the case of the 160 nm grating. Therefore, an evaluation of the scattering integrals is required to reveal the dominant detection path, especially in the case of the gratings with the two shortest periods.

The diffraction path in Eq. (S5-25) is given by:

$$J_{d-} \equiv \int_0^\infty \frac{\Delta\tilde{\varepsilon}_-(x_3)}{\varepsilon} e^{i\left(k_n + \sqrt{k_n^2 - k_g^2}\right)x_3} dx_3 = \varepsilon \int_0^\infty p'_{44}(x_3)\tilde{\eta}_1(x_3)\, e^{i\left(k_n + \sqrt{k_n^2 - k_g^2}\right)x_3} dx_3. \qquad \text{(S5-26)}$$

The layering in depth of the photo-elastic response, caused by the enhanced photo-elastic constants in the exciton layer and the (presumably weak) difference in the photoelasticities of GaAs and AlGaAs splits the integration in (S5-26) in several regions, where the integrals are the same. We are mostly interested in the signal from the GaAs excitonic layer, which is described by

$$J_{d-}(X) = \varepsilon p'_{44}(\mathrm{X}) \int_{h_f}^{h_r} \tilde{\eta}_1(x_3)\, e^{i\left(k_n + \sqrt{k_n^2 - k_g^2}\right)x_3} dx_3, \qquad \text{(S5-27)}$$

where $h_{f,r}$ denote the depth coordinates of the front and rare surfaces of the GaAs excitonic layer. The strain in the RSAW varies in depth as sum of two different exponents of different amplitudes [5]. We present the strain distribution in the form, $\tilde{\eta}(x_3) = a_m e^{-\alpha_m x_3}$, where $m$ = 1, 2 is a summation index. This form is valid both for $\tilde{\eta}_1$ and $\tilde{\eta}_3$, although with different constitutive parameters. Therefore, the integration in (S5-27) is straightforward:

$$J_{d-}(X) = \varepsilon p'_{44}(\mathrm{X}) a_m \frac{\left[e^{\left(ik_n + i\sqrt{k_n^2 - k_g^2} - \alpha_m\right)x_3}\right]_{h_f}^{h_r}}{i\left(k_n + \sqrt{k_n^2 - k_g^2}\right) - \alpha_m}. \qquad \text{(S5-28)}$$

The evaluation of the non-diffraction contribution in Eq. (S5-25) $J_{p+} \equiv \int_0^\infty \frac{\overline{\Delta\varepsilon_+}}{\varepsilon} e^{2ik_n x_3} dx_3 = \varepsilon \int_0^\infty \overline{p'_{12}\eta_3}\, e^{2ik_n x_3} dx_3$ is similar, leading to

$$J_{p+}(X) = \varepsilon p'_{12}(\mathrm{X}) a_m \frac{\left[e^{(2ik_n - \alpha_m)x_3}\right]_{h_f}^{h_r}}{2ik_n - \alpha_m}. \qquad \text{(S5-29)}$$

In Eqs. (S5-27) and (S5-29), the modification by the exciton of the refractive index of GaAs is negligible and, therefore $k_n \approx 2\pi n_{\mathrm{GaAs}}/\lambda_0$, where $n_{\mathrm{GaAs}} = \mathrm{Re}\, n \approx 3.67$ as determined in the main text. If it is required to evaluate the scattering from the other parts of the sample, it is sufficient in the derived formulas to choose the appropriate integration limits. The difference in the refractive indexes of GaAs and AlGaAs can be neglected.

## 6. Estimates of probe transmission and reflection by metal grating and their anisotropy

We consider the probe light to hit the sample surface at normal incidence. The transmission coefficient $t_{02}$ for its electric field component from air across the gold film (grating) of thickness $h$ into the GaAs substrate is described by the formula:

$$t_{02} = \frac{t_{01}t_{12}e^{-ik_1h}}{1+r_{01}r_{12}e^{-2ik_1h}},$$

where $t_{ab} = \frac{2n_a}{n_{a+}n_b}$, $r_{ab} = \frac{n_a-n_b}{n_{a+}n_b}$ denote the electric field transmission and reflectivity coefficients, the indices *a, b* equal *0* for air, *1* – for the gold film and *2* – for the GaAs substrate, and $k_1$ is the probe light wavevector. With this formula and the optical parameters of gold and GaAs at the probe wavelength $\lambda_0 \cong 820$ nm, $n_1 \cong 0.16 - i5.07$ [13], $n_2 \equiv n \cong 3.67 - i0.08$ [11], we estimate that the gold film of $h$ =10 nm diminishes both the transmission coefficient amplitude and phase by about 30%, i.e., $\frac{t_{02}(H)-t_{02}(0)}{t_{02}(0)} \approx -\left|\frac{\Delta t_{02}(h)}{t_{02}(0)}\right| e^{-i\left(\frac{\Delta\varphi_{02}(h)}{2\pi}\right)2\pi}$, where $\left|\frac{\Delta t_{02}(h)}{t_{02}(0)}\right| \approx \left|\frac{\Delta\varphi_{02}(h)}{2\pi}\right| \approx 0.3$.

For the estimates we assume that the optically thin gold film provides a localized, at $x_3 = 0$, rectangular modulation of the transmission coefficient with 50% pitch: $t_{02} = t_{02}(0)\left[1 + \left(\frac{t_{02}(h)-t_{02}(0)}{t_{02}(0)}\right)\right]$. We split the contribution from the grating, i.e., $\left(\frac{t_{02}(h)-t_{02}(0)}{t_{02}(0)}\right)$, into its averaged part and an oscillating part with zero average over a period, $t_{02} = t_{02}(0)\left[1 + \overline{\left(\frac{t_{02}(h)-t_{02}(0)}{t_{02}(0)}\right)} + \left(\frac{t_{02}(\widetilde{h)-t_{02}}(0)}{t_{02}(0)}\right)\right]$, where $\overline{\left(\frac{t_{02}(h)-t_{02}(0)}{t_{02}(0)}\right)} = \frac{1}{2}\left(\frac{t_{02}(h)-t_{02}(0)}{t_{02}(0)}\right)$, while the amplitude and phase in the oscillating part $\left(\frac{t_{02}(\widetilde{h)-t_{02}}(0)}{t_{02}(0)}\right)$ are $\frac{1}{2}\left|\frac{\Delta t_{02}(h)}{t_{02}(0)}\right|$ and $\frac{1}{2}\left|\frac{\Delta\varphi_{02}(h)}{2\pi}\right|$, respectively, with rectangular modulation. Therefore, relative to the averaged level of the transmission coefficient, the modulation of both amplitude and phase is at the level of 15%:

$$\left|\left(\frac{t_{02}(\widetilde{h)-t_{02}}(0)}{t_{02}(0)}\right)\right| \sim 0.15 \qquad \text{(S6-1)}$$

Moreover, when neglecting the higher spatial harmonics of the lateral modulation profile, we retain the modulation at the fundamental harmonic, which is $\pi$ times smaller in amplitude than the rectangular one:

$$t_{02} \cong t_{02}(0)\left[1 + \frac{1}{2\pi}\left|\frac{t_{02}(h)-t_{02}(0)}{t_{02}(0)}\right|\cos\left(\frac{2\pi}{d}x_1\right)\right] \cong t_{02}(0)\left[1 + \left(\frac{\widetilde{t_{02}}}{t_{02}}\right)\cos\left(\frac{2\pi}{d}x_1\right)\right].$$

Here the zero of the coordinate $x_1$ is chosen in the middle of the gold bar. This estimate suggests a modulation at the fundamental spatial period with an amplitude: $m \equiv \left(\frac{\widetilde{t_{02}}}{t_{02}}\right) \approx 0.05$. Note, that phase modulation is completely neglected in the above formulas.

The most general formulas for the detection process, derived in the Supplemental material, Section 5, indicate that the anisotropy of the optical properties of the metal grating opens additional opportunities for the detection of the SAWs, i.e., there potentially exist components of the SAW strain, which are undetectable by isotropic gratings (see Eq. (S5-22) from the Supplemental material, Section 5). Therefore, it is required to evaluate the modulation of the transmission coefficients in presence of optical anisotropy of the grating, which is dominantly caused by the effect of plasmon resonance.

Therefore, to generalize the above estimates, we assume that the transmission coefficient is modified additively by the presence of the grating and the presence of the plasmon resonant effect, where both contributions are rectangularly modulated with 50% pitch:

$$t = t_0 + \bar{t}_g + \bar{t}_p + \tilde{t}_g + \tilde{t}_p,$$

The amplitude of the oscillating terms is equal to the average contributions, $|\tilde{t}| = |\bar{t}|$, while the indexes "*g*" and "*p*" denote the isotropic contribution from the grating and the anisotropic contribution from its plasmon resonance, respectively. Therefore, evaluating the isotropic and anisotropic parts of the transmission coefficient gives:

$$t_+ \equiv \frac{t_2+t_1}{2} = t_0 + \bar{t}_g + \bar{t}_{p+} + \tilde{t}_g + \tilde{t}_{p+} \cong t_0\left[1 + \frac{\tilde{t}_g}{t_0} + \frac{\tilde{t}_{p+}}{t_0}\right] = t_0[1 + m_+],$$

$$t_- \equiv \frac{t_2-t_1}{2} = \bar{t}_{p-} + \tilde{t}_{p-} \cong t_0\left[\frac{\bar{t}_{p-}}{t_0} + \frac{\tilde{t}_{p-}}{t_0}\right] = t_0\left[\frac{\bar{t}_{p-}}{t_0} + m_-\right], \qquad \text{(S6-2)}$$

where the indexes *1* and *2* are used for the two orthogonal orientations of the electric field vector (along axis O$x_1$ and O$x_2$, respectively). The formulas above contain the modulation coefficients $m_\pm$, required in the detection Eq.(S5-22) of the Supplemental material, Section 5. Currently, we have an estimate only for a part of $m_+$, i.e., $\left|\frac{\tilde{t}_g}{t_0}\right| \sim 0.15$ (see Eq. (S6-1) above).

To estimate the other contributions to $m_\pm$, we derive $\frac{\bar{t}_{p-}}{t_0}$ from our experimental measurements of the optical reflectivity anisotropy in Fig. S6. In fact, the transmission and reflection coefficients for the embedded gold film (*1*), are related by the classic formula $t_{02}t_{20} = 1 + r_{02}r_{20}$, or $\frac{n_2}{n_0}(t_{02})^2 = 1 - (r_{02})^2$. Therefore, any small changes in the transmission and reflection coefficients are related by $\frac{n_2}{n_0}\left(\frac{\Delta t_{02}}{t_{02}}\right) \cong -\left(\frac{\Delta r_{02}}{r_{02}}\right)$. For the case of small changes due to optical anisotropy, the latter relation suggests $\frac{\bar{t}_{p-}}{t_0} \cong \frac{n_0}{n_2}\frac{\bar{r}_{p-}}{r_0}$, where the anisotropy of the reflection coefficients can be evaluated from the data in Fig. S6, as follows:

$$\frac{R_x}{R_y} \equiv \frac{R_1}{R_2} = \frac{\overline{r_1}^2}{\overline{r_2}^2} = \frac{(\overline{r_+}-\overline{r_-})^2}{(\overline{r_+}+\overline{r_-})^2} = \frac{\left(r_0+\overline{r_g}+\overline{r_{p+}}-\overline{r_{p-}}\right)^2}{\left(r_0+\overline{r_g}+\overline{r_{p+}}+\overline{r_{p-}}\right)^2} \cong \frac{\left(1-\frac{\overline{r_{p-}}}{r_0}\right)^2}{\left(1+\frac{\overline{r_{p-}}}{r_0}\right)^2} \cong 1 - 4\frac{\overline{r_{p-}}}{r_0},$$

for which we used the same principle of notations as in Eq. (S6-2) for the transmission coefficients.

Using the experimental relation $\frac{R_x}{R_y} = 1 + a$ with $a \leq 0.2$ we, first, estimate the anisotropy of the optical reflection coefficients, required in Eq. (S5-20) in the Supplemental material, Section 5: $\left|\left(\frac{r_-}{r_0}\right)\right| \equiv \left|\frac{\overline{r_{p-}}}{r_0}\right| \cong \frac{a}{4} \leq 0.05$. This provides the estimate of $\left|\frac{\bar{t}_{p-}}{t_0}\right| \cong \frac{n_0}{n_2}\left|\frac{\overline{r_{p-}}}{r_0}\right| \leq 0.015$ taking $n_0 = 1$ and $n_2 \cong 3.67$. The same estimate is valid for the modulated components $\left|\frac{\tilde{t}_{p+}}{t_0}\right| \sim \left|\frac{\tilde{t}_{p-}}{t_0}\right| \leq 0.015$, contributing to the modulation coefficients $m_\pm$. Therefore $m_+ \approx \frac{\tilde{t}_g}{t_0} \approx 0.15$ is controlled by the grating, while $m_- \approx \frac{\tilde{t}_{p-}}{t_0} \approx 0.015$ is controlled by the plasmon resonance, and $|m_-| \ll |m_+|$.

These inequalities are applied in the Supplemental material, Section 5 for the transition from Eq.(S5-22) to (S5-24) and further.

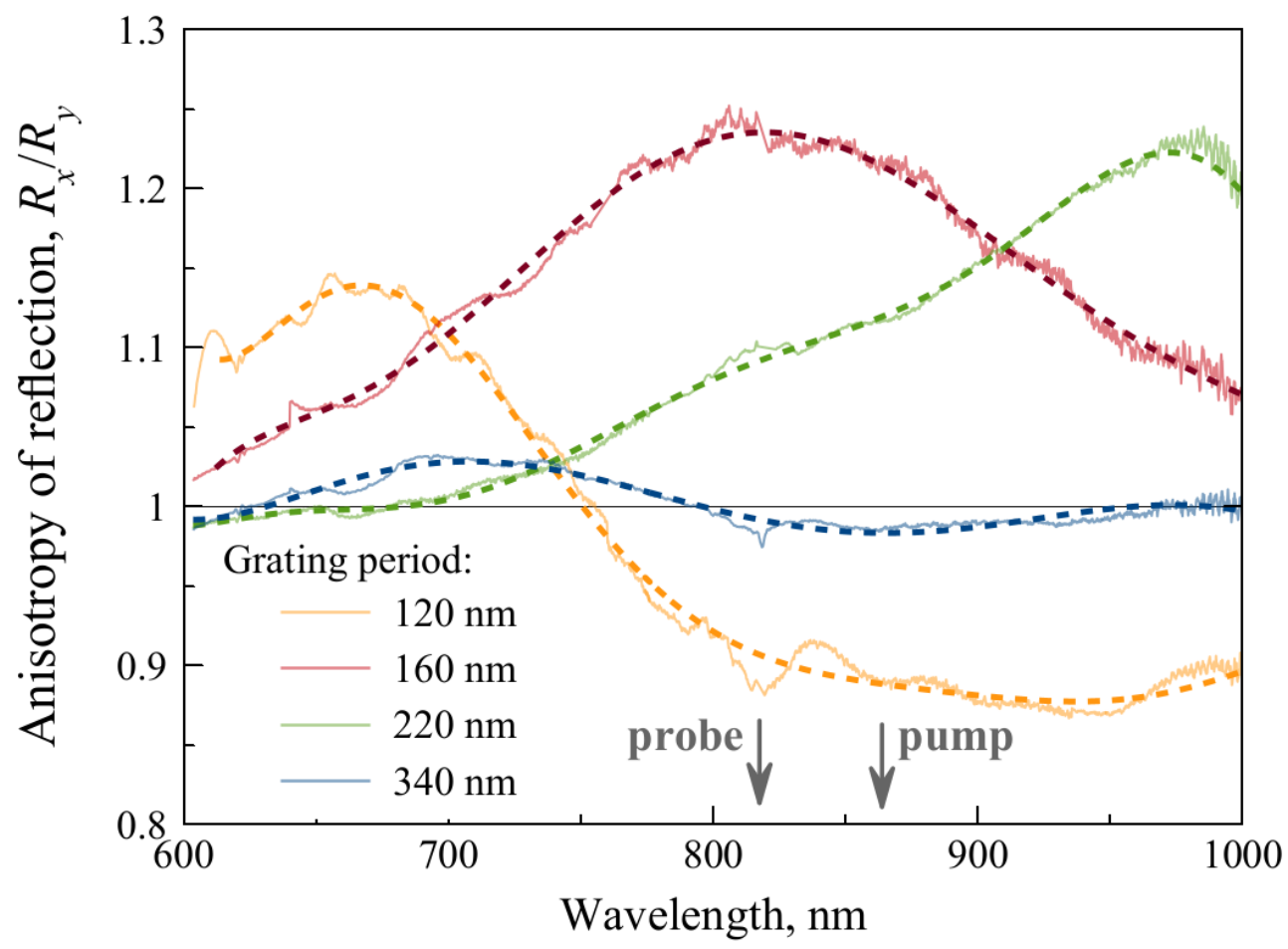

FIG. S6. Anisotropy of the reflection coefficients of the Au gratings with different periods on the AlGaAs/GaAs heterostructure, showing the ratio $R_x/R_y$ of the reflectivities for light linearly polarized perpendicular (along the O$x$ axis) and parallel (along the O$y$ axis) the Au stripes of the gratings. Features due to interference of light reflected from the heterointerfaces are subtracted. Arrows show the pump and probe wavelengths (865 and 818 nm, respectively).